\documentclass[usenatbib]{mn2e}
\usepackage[dvips]{graphicx}

\def\CN2{\mbox{$C_N^2 \ $}}

\def\CT2{\mbox{$C_T^2 \ $}}

\def\seeFA{\mbox{$\varepsilon_{FA} \ $}}

\def\sigmal2{\mbox{$\sigma ^{2}_{I} \ $}}

\title[Optical turbulence vertical distribution at Mt.Graham]{Optical turbulence vertical distribution with standard and high resolution at Mt. Graham}
\author[E. Masciadri et al.]{E. Masciadri,$^{1}$\thanks{E-mail:
     masciadri@arcetri.astro.it} J. Stoesz$^1$, S. Hagelin$^{1, 2}$, F. Lascaux$^1$  \\ $^1$INAF Osservatorio Astrofisico
di Arcetri, Largo Enrico Fermi 5, I-501 25 Florence, Italy\\
$^2$Uppsala Universitet, Department of Earth Sciences, Villav\"agen 16,
S-752 36 Uppsala, Sweden}

\begin{document}
\label{firstpage}
\date{Accepted 2009 ??? ??, Received 24 September 2009; in original form
2009 ??? ??}  
\pagerange{\pageref{firstpage}--\pageref{lastpage}}
\pubyear{2009}

\maketitle

\begin{abstract}
A characterization of the optical turbulence vertical distribution ($\CN2$ profiles) and all the main integrated astroclimatic parameters derived from the $\CN2$ and the wind speed profiles above the site of the Large Binocular Telescope (Mt. Graham, Arizona, US) is presented. The statistic includes measurements related to 43 nights done with a Generalized Scidar (GS) used in standard configuration with a vertical resolution $\Delta$H$\sim$1 km on the whole 20 km and with the new technique (HVR-GS) in the first kilometer. The latter achieves a resolution $\Delta$H$\sim$20-30 m in this region of the atmosphere. Measurements done in different periods of the year permit us to provide a seasonal variation analysis of the $\CN2$. A discretized distribution of $\CN2$ useful for the Ground Layer Adaptive Optics (GLAO) simulations is provided and a specific analysis for the LBT Laser Guide Star system ARGOS (running in GLAO configuration) case is done including the calculation of the {\it 'gray zones'} for J, H and K bands. Mt. Graham confirms to be an excellent site with median values of the seeing without dome contribution $\varepsilon$ = 0.72", the isoplanatic angle $\theta_{0}$ = 2.5" and the wavefront coherence time $\tau_{0}$ = 4.8 msec. We find that the optical turbulence vertical distribution decreases in a much sharper way than what has been believed so far in proximity of the ground above astronomical sites. We find that 50$\%$ of the whole turbulence develops in the first 80$\pm$15 m from the ground. We finally prove that the error in the normalization of the scintillation that has been recently put in evidence in the principle of the GS technique, affects these measurements with an absolutely negligible quantity (0.04"). 
\end{abstract}


\section{Introduction}
The Mt. Graham International Observatory (MGIO) is located on Mt. Graham (32$^{\circ}$42'05" N, 109$^{\circ}$53'31" W), Arizona (US) and hosts three telescopes: the Vatican Advanced Technological Telescope (VATT - D = 1.83 m) the Heinrich Hertz Submillimiter Telescope (SMT - D = 10 m) and the Large Binocular Telescope (LBT - two D = 8.4 m dishes that, when working in an  interferometric configuration, can achieve the resolution of a telescope with a 23 m pupil size). The study and characterization of the optical turbulence (OT) distribution in space and time is fundamental for the ground-based astronomy in the visible up to the near-infrared range to design adaptive optics systems and to optimize their performances. The vertical distribution of the OT (i.e. the $\CN2$ profiles) is the parameter from which all the integrated astroclimatic parameters derive from. 

In this paper we present a study based on measurements of the $\CN2$ profiles related to 43 nights and obtained with a Generalized Scidar (GS) placed at the focus of the VATT on the Mt.Graham summit, around 250 m from the LBT. Figure 4 in Egner \& Masciadri (2007) shows the relative position of the two telescopes. It is worth to note that the primary mirror of the LBT is located $\sim$ 35 m above the dome of the VATT.  

The scientific motivations for such a long-term site testing campaign are: \newline
{\bf (1)} To collect an as rich as possible statistical sample of optical turbulence (OT) vertical distribution ($\CN2$ profiles) to be compared with simulations obtained with the atmospherical mesoscale model Meso-Nh with the aim to validate the model above Mt. Graham. This is a key milestone for the ForOT project\footnote{$http://forot.arcetri.astro.it$} whose final goal is to predict the optical turbulence above astronomical sites (Masciadri 2006). The measurements from a vertical profiler such as a Generalized Scidar are crucial for the validation of such a kind of models. It is our interest to collect a heterogeneous sample of measurements taken in different periods of the year and different turbulence conditions in a way to better control the model behaviors under different conditions and to better validate the model itself. The atmospherical models has been used for the first time to reconstruct and characterize the $\CN2$ profiles by Masciadri et al. (1999a, 1999b). Since then many progresses have been achieved by our group: the model has been applied to different astronomical sites in a simple monomodel configuration (Masciadri et al. 2001, Masciadri \& Garfias 2001) and more recently in a grid-nesting configuration (Lascaux et al. 2009), a new calibration technique has been proposed (Masciadri \& Jabouille 2001) and statistically validated (Masciadri et al. 2004) and the first application of the Meso-Nh model as a tool of turbulence characterization has been presented (Masciadri \& Egner 2006). However, so far we always could access to GS measurements concentrated in a well defined period of time. This series of site testing campaigns with a GS at Mt. Graham aimed to comply with the necessity to diversify the measurements sample. \newline
{\bf (2)} To provide a characterization of all the most important integrated astroclimatic parameters above the site of the Large Binocular Telescope (LBT) and verify if any evident changes are observed with respect to previous results obtained on a three times smaller sample  (Egner et al. 2007). This study can be, therefore, important to confirm/reject/refine those conclusions. We remind that the GS is a manually operated instrument and the statistical richness of GS measurements can not be compared to that of monitors such as the DIMM and the MASS that are routinely run above many observatories. However, these monitors can provide only integrated values (as is the case for the DIMM) or turbulence distribution with very low vertical resolution $\Delta$H $\sim$ h/2 (as it is the case for the MASS). The GS remains therefore  a unique instrument for a detailed analysis of the vertical distribution of the OT in the whole troposphere. \newline 
{\bf (3)} To provide an as rich as possible statistical sample of the high resolution vertical distribution ($\Delta$H = 200-250 m and 25-30 m) of the optical turbulence in the first hundreds of meters up to 1 km to support the feasibility studies of new generation instruments for the LBT, such as the LBT Laser Guide Stars system ARGOS (Rabien et al. 2008), that, in its first baseline, is planned to work with a GLAO\footnote{Ground Layer Adaptive Optics (GLAO)} configuration. The GLAO efficiency, indeed, strongly depends on the turbulence distribution and strength near the ground. The study we intend to perform can be achieved thanks to a new technique that has been recently proposed called High-Vertical Resolution Generalized Scidar (HVR-GS - Egner \& Masciadri 2007) that aims to reconstruct the optical turbulence vertical distribution in the first kilometer above the ground with a resolution 4 up to 10 times higher than what has been done so far with standard vertical profilers such as the GS (typically $\Delta$H $\sim$ 1km). In Egner \& Masciadri (2007) the validity of this technique has been proved and it is now our intention to characterize the turbulence distribution in statistical terms, to verify how the turbulence decreases in the first hundreds of meters above the ground and provide inputs to test the ARGOS performances.

In Section \ref{instr} we briefly review the principle of the GS and HVR-GS techniques. In Section \ref{campaign} we present the site testing campaign data-set that has been analyzed and presented in a preliminary form in Stoesz et al. (2008). In Section \ref{astro_integ} we present an exhaustive statistical analysis of all the most important integrated astroclimatic parameters derived from the $\CN2$ profiles and characterizing Mt. Graham. In Section \ref{composit} we quantitatively discuss the turbulence distribution of the OT in the whole troposphere in application to the Adaptive Optics. We provide a composite profiles distribution on the whole troposphere with the vertical resolution required for the characterization of the LBT-LGS system running in GLAO configuration (ARGOS) i.e. ($\Delta$H $\sim$ 200 m in the first kilometer and $\Delta$H $\sim$ 1 km above 1 km). In Section \ref{cn2} we characterize the $\CN2$ profiles at standard and high-vertical resolution and their seasonal variation. Section \ref{concl} summarizes the conclusions. In Annex we briefly show that the error in the normalization of the scintillation of the GS technique, put in evidence by Johnston et al. (2002) and Avila \& Cuevas (2009), produces effects absolutely negligible on these measurements. From a general point of view, we will show that GS measurements obtained with a pupil size D $\ge$ 1.5 m and a binary separation $\theta$ $\le$ 8 arcsec, are affected by this error for less than a few hundredths of an arcsec.

\section{Instruments: GS and HVR-GS}
\label{instr}

Two instruments have been used for this study: the GS and the HVR-GS. We used the GS as developed by McKenna et al. (2003). 
The Scidar technique has been originally proposed by Rocca et al. (1974) and Vernin \& Azouit (1983) and relies on the analysis of the scintillation images generated by a binary in the pupil plane of a telescope. The Scidar technique (called Classic Scidar) is insensitive to the turbulence near the ground. More recently Fuchs et al. (1998) proposed a generalized version of the Scidar (called Generalized Scidar) in which the detector is virtually conjugated below the ground permitting to extend the measurements range to the whole atmosphere ($\sim$ 20-25 km). The GS principle has been later put in practice by several authors above different astronomical sites (Avila et al. 1997, 2004, Klueckers et al. 1998, McKenna et al. 2003, Fuensalida et al. 2004, Egner et al. 2007, Garcia-Lorenzo et al. 2009). The GS is based on the observation of binaries with a typical separation $\theta$ within (3-10) arcsec, the binary magnitude m$_{1}$,m$_{2}$ $\le$ 5 mag and $\Delta$(m$_{1,2}$) $\le$ 1 mag.
When two plane wavefronts propagating from a binary meet a turbulence layer located at a height $h$ from the ground, they produce on the detector plan, optically placed below the ground at about h$_{gs}$ $= $ 3 km, two scintillation maps made by a set of characteristic shadows appearing in couple separated by a distance $d$ that is geometrically related to the position of the turbulence layer as $d$ $=$ $\theta$$\cdot$ ($h$ + h$_{gs}$). The calculation of the auto-correlation (AC) of the scintillation map produces the so called 'triplet'. The central peak is located in the centre of the AC frame, the lateral peaks are located at a symmetric distance $d$ from the centre. The amplitude of the later peaks is proportional to the strength of the turbulence of the layer located at the height $h$ weighted by the scintillation that such a layer produces on the detector.

In a multilayer atmosphere different turbulent layers produce triplets with lateral peaks located at a different distance $d$$^{*}$ from the centre of the AC frame. To monitor the whole troposphere the pupil size needs to be large enough (D $\ge$ 1.5 m) to avoid that one of the shadows that form a couple (or both of them) falls outside the pupil of the telescope. In the AC frame the triplets are all placed along the direction of conjunction of the binary.  To retrieve the $\CN2$ profile, the central peak of the triplet in which some sources of noise such as the photon noise are present, is firstly eliminated. Finally the $\CN2$ is obtained inverting the Fredholm equation that, in the LBT-GS, is done using the conjugated gradient (Egner et al. (2007)) algorithm. The vertical resolution $\Delta H(h)$ of the GS technique depends on our ability in discriminating two different later peaks and it is proportional to the Fresnel Zone (FZ) (Vernin \& Azouit, 1983):
\begin{equation}
\Delta H(h) = \frac{{0.78 \cdot \sqrt {\lambda | h - h_{gs}|} }} {\theta }  = \frac{{0.78 \cdot FZ}} {\theta }  
\label{res}
\end{equation}
For typical values of the observable binaries with a standard GS the typical vertical resolution is $\Delta H(0)$ $\sim$ 1 km. 

The HVR-GS technique has been introduced recently (Egner \& Masciadri, 2007) and aims to measure the $\CN2$ profiles with a high vertical resolution ($\Delta$H $\sim$ 25-30 m) in the first kilometer above the ground. We briefly summarize here the main concepts on which this technique is based on. If we abandon the idea to monitor the whole 20 km and we limit our attention to the first kilometer we can easily increase the vertical resolution of a GS up to a factor 4 using the standard GS technique based on the calculation of the autocorrelation (AC) obtained from the scintillation maps of binary stars having a separation $\theta$ around four times larger than the typical separation used for the standard GS technique. This is not enough, however, to achieve resolution of the order of 25-30 m because we are fundamentally limited by the Fresnel Zone size. However, if we use simultaneously the autocorrelation (AC) and the cross-correlation maps (CC) taken with 20-40 msec time lag, the triplets in the CC frames are not aligned anymore on the same direction identified by the binary separation but the central peak of each triplet is located on the direction of the wind speed of each turbulent layer distributed in the troposphere and the distance from the centre of the CC frame is proportional to the wind speed of the same turbulent layer. Under this assumption it is possible to prove (Egner \& Masciadri 2007) that the vertical resolution $\Delta h_{pix}$ is finally limited by the accuracy with which we can estimate the position of the lateral peaks in the CC frames that is smaller than the pixel size projected on the pupil ($\Delta$ x$_{pix}$) and it is equal to:
\begin{equation}
\Delta h_{pix}  = 0.56 \frac{{\Delta x_{pix} }} {\theta }
\end{equation}

For $\theta$ $=$ 30 arcsec and $\Delta$ x$_{pix}$ $=$ 7 mm we retrieve a typical vertical resolution $\Delta$ h$_{pix}$ $\sim$  25-30 m. In conclusion, the main HVR-GS concept consists in: {\bf (1)} taking a wide binary of the order of 30-35 arcsec monitoring the first kilometer and {\bf (2)} treating simultaneously the AC and CC frames. From a practical point of view, the $\CN2$ retrieved from the AC frames is characterized by an energy that is redistributed in a set of thinner layers within the first kilometer whose vertical resolution is of the order of 25 m. The AC frames are fundamental to re-normalize the total energy in the first kilometer. The final  high vertical resolution $\CN2$ profile is retrieved from:
\begin{equation}
\int\limits_{ - \Delta h_{\max } (0)/2}^{h_{max}} {C_{N,AC}^2 } (h) \cdot dh = f_{scale}  \cdot \sum\limits_i {C_{N,CC}^2 } (h_i )
\end{equation}
where h$_{max}$ is the height of the highest detected layer, the inferior limit of the integral $\Delta$h$_{max }$(0)/2 takes into account the vertical resolution at h=0. f$_{scale}$ is a factor that takes into account the decorrelation of the central peak of the triplets having a V $>$ 0 $\pm$ $\Delta$V ($\Delta$V = 0.2-0.8 m$\cdot$s$^{-1}$) with respect to the zero velocity triplet. It has been measured (Egner \& Masciadri, 2007) that f$_{scale}$ in the range $\Delta$H $=$ 1 km can be considered the same for all the thin turbulent layers. We note that, using the HVR-GS we have to consider a rate of rejection of frames because of the wind fluctuations that can produce a spreading of the central peak of the triplets. The rate of rejection can be more or less conservative, depending on the constraints that the user wishes to introduce in this analysis. Anyway, in Egner \& Masciadri (2007) it has been proved that no bias in the measurements is introduced if we reduce the statistical sample. This can be explained with the fact that the conditions that facilitate the spreading of the central peaks are not necessarily correlated to typical bad or good seeing. We refer the reader to that paper for further details. We highlight that two slightly different approaches have been proposed recently in the literature to increase the vertical resolution near the ground: the HVR-GS (Egner \& Masciadri 2007) and LOLAS (Avila et al. 2008, Avila \& Chun, 2004). The main difference from the point of view of the principle is the following: the HVR-GS uses a known instrument (the Generalized Scidar) but with a new technique. LOLAS uses a new instrument  but with the same technique of the Generalized Scidar. The readers can find in the corresponding papers the details (and technicalities) related to both.

\section{Site testing campaigns: data-set statistic}
\label{campaign}

\begin{table}
\begin{tabular}{cc}
\hline
Observing Runs & Nights \\
\hline
27 April 2005 & 1 \\
19-24 May 2005 & 6 \\
6-15 December 2005 & 5 \\
27 May 2007 - 3 June 2007  & 8 \\
16-28 October 2007 & 13 \\
23 February 2008 - 3 March 2008 & 10 \\
\hline
\end{tabular}
\caption{Observing runs at Mt. Graham.}
\label{tab_obs_run}
\end{table}

\begin{table}
{\begin{tabular}{@{}ccccc@{}}
\hline
Sample & Nights & Measurements & Hours & Resolution\\
\hline
'GS' & 43 & 16657 &  163 &  $\Delta$H(0) $\sim$ 1km\\
'WB' & 15 & 3659 &  6.2 &  $\Delta$H(0) $\sim$ 200 m\\
'HVR-GS' & 15 & 2812 &  5.1 &  $\Delta$H(0) $\sim$ 25 m\\
\hline
\end{tabular}}
\caption{Classification of the GS campaign measurements. {\bf 'GS':} $\CN2$ profiles retrieved from the standard GS.  {\bf 'WB':} $\CN2$ profiles retrieved from the AC frames of the GS measurements obtained with wide-binaries. {\bf 'HVR-GS':} $\CN2$ profiles retrieved from the wide-binaries auto-correlation (AC) and cross-correlation (CC) frames following the technique described in Egner \& Masciadri (2007).}
\label{tab_sample}
\end{table}

\begin{table}
{\begin{tabular}{@{}cccccc@{}}
\hline
Name & $\alpha_{J2000}$ & $\delta_{J2000}$ & m$_{1}$ & m$_{2}$ & $\theta$ \\
    & & &         (mag) & (mag) & (arcsec) \\
\hline
$\gamma$ Ari & 01 54 & + 19 17 &  4.5 &  4.6 & 7.6  \\
Castor & 07 35 & + 31 53 &  1.9 &  3.0 & 4.4  \\
$\gamma$ Leo & 10 20 & + 19 50 &  2.4 &  3.6 & 4.7  \\
$\pi$ Boo & 14 41 & + 16 25 &  4.9 &  5.8 & 5.5  \\
$\delta$ Ser & 15 34 & + 10 32 &  4.2 &  5.2 & 4.1  \\
95 Her & 18 02 & + 21 36 &  4.9 &  5.2 & 6.5  \\
\hline
$\beta$ Cyg & 19 31 & + 27 58 &  3.2 &  4.7 & 35.3  \\
\hline
\end{tabular}}
\caption{Binary stars observed with the GS and HVR-GS.}
\label{tab_stars}
\end{table}

We collected and analyzed so far observations related to 43 nights (Table \ref{tab_obs_run}). As already noted the GS is manually operated therefore the site testing campaigns have been scheduled respecting the periods in which the VATT was shut down (July and August) and trying to cover as many different periods of the year. Table \ref{tab_sample} shows the code used to identify the typology of the data-set: 'GS' indicates the standard GS measurements extended on $\sim$ 20 km (43 nights), 'WB' indicates the $\CN2$ retrieved from the autocorrelation (AC) frames associated to wide binaries ($\Delta$H $\sim$ 200-250 m) (15 nights) and 'HVR-GS' the $\CN2$ obtained with high vertical resolution ($\Delta$H $\sim$ 25-30 m) (15 nights) in which both the autocorrelation and cross-correlation frames have been treated. The samples of the three categories are differently rich because we started to use the HVR-GS more recently and this new technique has a higher rate of measurements rejection. We note that, to avoid biases in the estimates in the HVR-GS, it has been decided to discard from the statistic all doubtful cases characterized by the presence of clouds or cirrus. A method that we called {\it 'normalization'} (and it will be described later) has been applied to the sample of high resolution measurements (15 nights). Among other advantages it permits us to provide a turbulence budget representative of the whole sample of 43 nights. Basically, the morphology of the turbulence energy distribution (shape of the $\CN2$ versus the height) is retrieved from the observation of wide-binaries for 15 nights and with the {\it 'normalization'} procedure the turbulence energy of the first kilometer detected with the standard GS is redistributed in thin turbulent layers according to the profile morphology reconstructed with the CC frames.
The selected binary stars for the standard GS and HVR-GS techniques are reported in Table \ref{tab_stars}.
They are substantially the same indicated in Egner et al. (2007) and Egner \& Masciadri (2007) but we eliminated from the sample those binaries with magnitude larger than 6 mag (i.e. 118 Tau has been discarded) to avoid potential biases due to a too weak intensity. 

\section{Integrated Astroclimatic parameters}
\label{astro_integ}

The seeing, the isoplanatic angle $\theta_{0}$, the wavefront coherence time $\tau_{0}$ and the equivalent velocity V$_{0}$ are defined as:

\begin{equation}
r_0  = \left[ {0.423 \cdot \left( {\frac{{2\pi }}
{\lambda }} \right)^2  \cdot \int\limits_0^\infty  {C_N^2 (h)dh} } \right]^{ - 3/5} 
\end{equation}

\begin{equation}
\varepsilon  = 0.98\frac{\lambda }
{{r_0 }}
\end{equation}

\begin{equation}
\theta _0  = 0.057 \cdot \lambda ^{6/5}  \cdot \left[ {\int\limits_0^\infty  {h^{5/3} C_N^2 (h)dh} } \right]^{ - 3/5} 
\end{equation}

\begin{equation}
V_0  = \left[ {\frac{{\int\limits_0^\infty  {V(h)^{5/3} C_N^2 (h)dh} }}
{{\int\limits_0^\infty  {C_N^2 (h)dh} }}} \right]^{3/5} 
\end{equation}

\begin{equation}
\tau _0  = 0.31\frac{{r_0 }}
{{V_0 }}
\label{tau0}
\end{equation}

\begin{figure*}
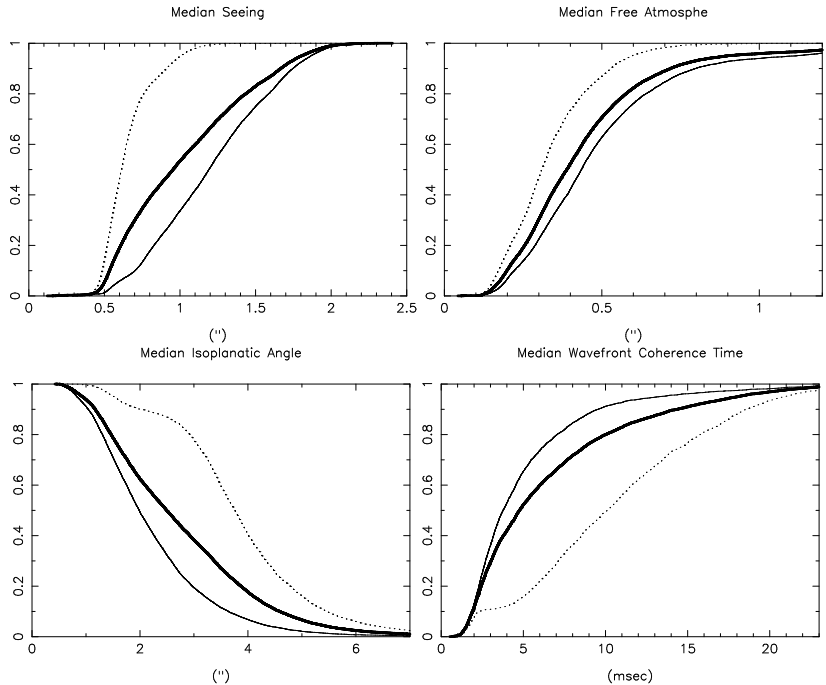

\centering
\includegraphics[width=4.5cm,angle=-90]{masciadri_fig1a}
\includegraphics[width=4.5cm,angle=-90]{masciadri_fig1b}

\includegraphics[width=4.5cm,angle=-90]{masciadri_fig1c}
\includegraphics[width=4.5cm,angle=-90]{masciadri_fig1d}
\caption{Cumulative distributions of four integrated astroclimatic parameters. Top-left: the total seeing (including the dome contribution). Top-right: the seeing in the free atmosphere (h $>$ 1 km). Bottom-left: the isoplanatic angle ($\theta_{0}$). Bottom-right: the wavefront coherence time ($\tau_{0}$). Thick lines: the whole sample. Dotted lines: summer time. Thin lines: winter time.
\label{astroclim_seas}} 
\end{figure*}

\begin{table*}
{\begin{tabular}{@{}cccccccccc@{}}
\hline
&Total &&& Summer & && Winter  && \\
\hline
Parameter & 25$^{th}$ & 50$^{th}$ & 75$^{th}$& 25$^{th}$ & 50$^{th}$ & 75$^{th}$& 25$^{th}$ & 50$^{th}$ & 75$^{th}$\\
\hline
$\varepsilon$ (arcsec)& 0.65 & 0.95 & 1.34 &0.53 &0.61 &0.72 &0.89 &1.19 &1.50  \\
$\theta_{0}$ (arcsec) & 1.6  & 2.5  &  3.6 &3.1 & 3.8 & 4.5& 1.4& 2.0& 2.7 \\
$\tau_{0}$ (msec) & 2.7 & 4.8 & 8.7& 6.4& 10.1&14.6 &2.5 &3.8 & 6.2 \\
V$_{0}$ (ms$^{-1}$) & 5.1 & 7.2& 9.3 &3.7 &5.1 &7.8 &5.9 &7.7 & 9.6 \\
\hline
\end{tabular}}
\caption{Median, first and third quartiles values of the main integrated astroclimatic parameters above Mt. Graham (43 nights): seeing in the total troposphere, isoplanatic angle, wavefront coherence time, integrated equivalent wind speed.}
\label{tab_integ_astro}
\end{table*}

Figures \ref{astroclim_seas} shows the cumulative distribution of the astroclimatic parameters (seeing, seeing in the free atmosphere calculated for h $>$ 1 km, isoplanatic angle and wavefront coherence time) calculated for the total 43 nights and for the (April-September) and (October-March) periods that we will simply call summer and winter\footnote{As shown in Table \ref{tab_obs_run}, measurements do not cover all the months of the year with an exact uniform distribution. For example there are no measurements in the July-September period. However we are interested in putting in evidence macroscopic differences between the two extreme seasons and we observed that the morphology of the average $\CN2$ profile and the turbulence strength was substantially the same including or not the data related to the October month. We therefore decided for this division instead of a more reductive (April-June) and (December-March) division that would have imply to study a less statistically representative sample.}. 
Table \ref{tab_integ_astro} summarizes the median (50$^{th}$), first (25$^{th}$) and third (75$^{th}$) quartiles for the three main integrated astroclimatic parameters calculated for the following groups: the whole sample of 43 nights, the summer and the winter time. 

A composite wind speed profile has been used to calculate the median wavefront coherence time $\tau_{0}$. Below 2 km the wind speed retrieved from the GS has been used; above 2 km the wind speed profile retrieved from the European Centre for Medium-Range Weather Forecast (ECMWF) analyses extracted in the nearest grid point (32.75$^{\circ}$N, 110.00$^{\circ}$W)\footnote{It is worth to highlight that the analyses from the ECMWF are available, at present, with a horizontal resolution of 0.25 degrees. Calculations done in Egner et al. (2007) have been obtained with data extracted at latitude (33$^{\circ}$N) and a horizontal resolution of 0.5 degrees.} to the Mt. Graham ($\sim$ 11.5 km northwest of the summit). Due to the fact that the wind speed vertical profiles retrieved from the ECMWF analyses are calculated at the synoptic hours (00:00, 06:00, 12:00, 18:00) a temporal interpolation of adjacent synoptic hours wind speed have been performed to better represent the wind speed in the local temporal range in which $\CN2$ measurements have been done. It has been already shown (Egner et al. (2007)) that this is the best method to treat the wind speed to calculate a reliable $\tau_{0}$. The ECMWF analyses do not well reconstruct the orographic effects produced on the atmospheric flow at the top of the summit. Moreover, the ECMWF grid points spaced by 0.25 degrees correspond to locations with lower altitudes than the summit altitude H$_{0}$ and, as a consequence, the ECMWF wind speed calculated at an height equal to H$_{0}$ is, in general, larger than the wind speed measured on the summit. On the other side, measurements of the wind speed with GS implies a great number of rejected frames and it is frequently difficult to retrieve a profile extended on the whole troposphere. The composition method for the wind speed revealed, therefore, to be the best solution for the calculation of $\tau_{0}$ (Egner et al. 2007). 

Looking at Fig.\ref{astroclim_seas} and Table \ref{tab_integ_astro} a clear seasonal variation appears evident for all the integrated astroclimatic parameters. In May 2005 and May 2007, in both years, 10-15 days of extremely good seeing and large isoplanatic angle  features  has been observed. Such a long time of extremely good conditions in the same period of the year indicate that this is, highly probably, among the best periods of the year for turbulence conditions at Mt. Graham. This result is coherent with the typical weather conditions at synoptic scale in this region and in this season that are characterized by weaker wind speed in the high troposphere. A low probability to trigger optical turbulence in the high atmosphere confirmed by a weaker $\CN2$ strength (see Section \ref{cn2}), is coherent with a large median $\theta_{0}$ and a small median seeing in the free atmosphere (Fig. \ref{astroclim_seas}) in the summer seasons. If we also take into account the typical weaker equivalent wind speed (Eq.\ref{tab_integ_astro}) in summer with respect to winter time (see Table \ref{tab_integ_astro} and Section \ref{cn2}) we can explain the typical larger value of the median $\tau_{0}$ in this season (see Eq.\ref{tau0}). 

\begin{figure}
\centering
\includegraphics[width=6 cm,angle=-90]{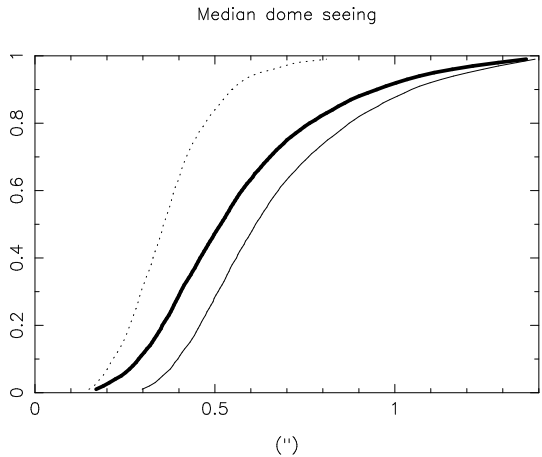}
\caption{Cumulative distribution of the dome seeing for all of the 43 nights (thick line), the summer (dotted line) and winter (thin line) period. 
\label{dome_see}} 
\end{figure}

Finally, to quantify the contribution of the seeing provided only by the atmosphere, the dome seeing ($\varepsilon_{d}$), calculated with the method described in Egner et al. (2007) and Avila et al. (2001), has been subtracted from the total seeing ($\varepsilon$). The method consists in discriminating the triplets located at the ground (H=H(0)$\pm$$\Delta$H) with a velocity V = 0 $\pm$ $\Delta$V from those located at the same height and having a velocity different from zero (V $>$ $\Delta$V). The velocity resolution $\Delta$V of our system is 0.8 to 0.2 $m/s$ per pixel with time lags within the range (10 to 40 $ms$).
Figure \ref{dome_see} shows the cumulative distribution of the dome seeing calculated for the whole sample and the two seasons. The median value of the 'dome seeing' is $\varepsilon_{d}$$=$ 0.52 arcsec (Fig.\ref{dome_see}). For the first time it has been observed an interesting seasonal trend in the dome contribution that certainly deserves a carefully investigation in the future. {\bf Knowing that the median seeing in the whole atmosphere (included the dome seeing) is $\varepsilon$$=$0.95 arcsec and that $\varepsilon_{d}$$=$ 0.52 arcsec, it follows that the median seeing related to the whole atmosphere without the dome contribution for the richest statistic we collected so far (43 nights) is $\varepsilon_{tot}$$=$ 0.72 arcsec}.\newline

\begin{table}
{\begin{tabular}{@{}ccc@{}}
\hline
Parameter &Egner et al. (2007) & Masciadri et al. (2009) \\
\hline
$\varepsilon$ included dome & 0.80 & 0.95 \\
$\varepsilon$ without dome & 0.68 & 0.72 \\
$\theta_{0}$ & 2.71 &  2.5 \\
$\tau_{0}$ &  3.6 & 4.8  \\
\hline
\end{tabular}}
\caption{Comparison of values obtained in the study of Egner et al. (2007 ) - 16 nights and Masciadri et al. (2009) - 43 nights. Seeing (with and without dome contribution) and $\theta_{0}$ are in arcsec, the $\tau_{0}$ in msec.}
\label{mascia_vs_egner}
\end{table}

Table \ref{mascia_vs_egner} reports the difference in the median estimates of the principal integrated astroclimatic parameters obtained above Mt. Graham with samples that are differently rich by a factor $\sim$ 3. No major differences can be highlighted with exception of the $\tau_{0}$ that appears larger of about 1 msec with respect to the paper of Egner et al. (2007).

To complete the analysis of the integrated astroclimatic parameters it is worth to remind that, in the context of GLAO systems (Tokovinin, 2004), the atmosphere can be divided in three vertical slabs: a region near the ground (h $<$ H$_{min}$) in which the turbulence is totally corrected, an intermedium region (H$_{min}$ $<$ h $<$ H$_{max}$), called {\it Ôgray zoneÕ} in which the turbulence is partially corrected and a region that covers the rest of the troposphere (h $>$ H$_{max}$) in which the turbulence is not corrected any more. The values of H$_{min}$ and H$_{max}$ depend on the wavelength, the field of view, the turbulence conditions, the pitch size of the adaptive optics system.
A more detailed discussion on that will be done later but it is, of course, evident that it can be very useful for the GLAO applications {\bf (1)} to know the budget of turbulence energy developed in the high part of the atmosphere i.e. how much turbulence remains not corrected by a GLAO system and {\bf (2)} to better estimate the size of the gray zone and the vertical distribution of the turbulence inside the gray zone. The gain of a GLAO system depends, indeed, mainly on these two issues. From the standard GS measurements it is trivial to calculate the cumulative distribution of $\seeFA$ for h $>$ 1 km (Figure \ref{astroclim_seas}). The same calculation is done for the whole year, the summer and winter time. We calculated a median value $\seeFA$ =  0.39 arcsec for h $>$ 1 km, ($\varepsilon_{FA,sum}$ =  0.31 arcsec, $\varepsilon_{FA,win}$ =  0.44 arcsec). For those cases in which H$_{max}$ = 1 km the $\seeFA$ represents the portion of turbulence that is not corrected at all. In the next session we will deal about the portion of turbulence developed above X m, where X $<$ 1km extremely important for GLAO applications.

\section{Composite profiles for Adaptive Optics applications}
\label{composit}

The {\it 'composite profiles'} are used to represent efficiently and in statistical terms the vertical distribution of the optical turbulence in a discretized number of layers particularly suitable for AO simulations. The method is commonly used by many authors in the field (Tokovinin \& Travouillon 2006, Egner et al. 2007, Stoesz et al. 2008, Masciadri et al. 2009, Chun et al. 2009) and it consists in identifying a finite number of vertical slabs covering the whole troposphere ($\sim$ 20 km) with their correspondent value of seeing ($\Delta$$\varepsilon$$_{i}$ or $\Delta$J$_{i}$) so that the total turbulence integrated on the troposphere is conserved. J is defined as:

\begin{equation}
J = \int\limits_0^\infty  {C_N^2 (h) dh} 
\end{equation}

and it is related to the seeing as:

\begin{equation}
J = 9 \cdot 10^{-11} \cdot \lambda ^{1/3}  \cdot \varepsilon ^{5/3} 
\end{equation}

where $\lambda$ is expressed in meters, $\varepsilon$ in arcsec and J in m$^{1/3}$. The turbulence in the free atmosphere (h $>$ 1 km) and in the boundary layer (h $\le$ 1 km) is treated in an independent way to permit to study different combinations of probabilities for the OT vertical distribution. 
Table \ref{comp_h_gt_1km} reports the $\Delta$J$_{i}$ values in the range h $>$ 1 km calculated at different heights and obtained from the $\CN2$ associated to the r$_{0}$ related to the 20-30 $\%$ of its cumulative distribution ('good' case), to the 45-55 $\%$ ('typical' case) and to the 70-80 $\%$ ('bad' case). Measurements from the sample 'GS' (see Table \ref{tab_sample}) are used. 

The $\CN2$ profiles retrieved from the 'WB' sample characterized by a $\Delta$h $\sim$ 200 m near the ground (H $<$ 1 km) have a suitable vertical resolution to calculate the composite profiles in this vertical range for applications to ARGOS (field of view $\theta$ = 4 arcmin) because the turbulence developed below H$_{min}$ = $\Delta$X/(2$\cdot$$\theta$) $\sim$ 200 m ($\Delta$X = 0.5 m is the pitch size i.e. the projection of the actuator of the deformable mirror on the pupil of the telescope) is resolved by the instrument. Basically we do not need a higher vertical resolution for this application. We therefore first calculated a similar distribution of Table \ref{comp_h_gt_1km} for h $\le$ 1 km and obtain a temporary table that we call table T similar to Table \ref{comp_h_gt_1km}. 
As already anticipated in Stoesz et al. (2008) and Masciadri et al. (2009), to take into account the different statistic obtained with the standard 'GS' and wide binary sample 'WB' and to take into account the quantitative information of the turbulence present in the first kilometer provided by the whole sample of 43 nights, each number of this temporary table T has to be multiplied by the correction factor f$_{gl}$:

\begin{equation}
f_{gl}  = \left( {\frac{{\varepsilon _0 }}
{{\varepsilon _{0'} }}} \right)^{5/3}  
\label{fgl}
\end{equation}

reported in Table \ref{fgl_factor} to finally obtain the composite distribution for h $\le$ 1 km (Table \ref{comp_h_lt_1km_corr}). The numerator of f$_{gl}$ is the seeing measured in the first kilometer from the ground from the 'GS' sample. The denominator is the seeing measured in the first kilometer from the ground from the 'WB' sample. Table \ref{comp_h_lt_1km_corr} reports the composite profiles for h $<$ 1 km equivalent to 43 nights. It includes the dome contribution frequently preferable for AO simulations. In the last row the J values in the case in which the dome contribution is subtracted are reported\footnote{We note that the median seeing of Table \ref{tab_integ_astro} is not exactly the same as the seeing retrieved by the composite profiles because the former treats a set of individual seeing values that are proportional to J$^{3/5}$.}. The multiplication by the f$_{gl}$ factor (method that we call {\it normalization}) offers the great advantage to retrieve the spatial distribution of the turbulence in the first kilometer using the HVR-GS technique (or the 'WB' as is this case) and to use the quantitative turbulence energetic budget of data extracted from the standard GS technique that can be done on a richer statistical sample (43 nights). We overcome therefore the intrinsic limitation of a high number of rejected frames typical of the cross-correlation technique. This method also is completely insensitive to any bias potentially introduced by the normalization of the autocovariance of the scintillation maps of the GS put in evidence by Johnston et al. (2002) and Avila \& Cuevas (2009) (see Annex B for demonstration).
The median value of the composite profiles (Table \ref{comp_h_gt_1km} and Table \ref{comp_h_lt_1km_corr} central columns) permits us to calculate the percentage of the turbulence developed above different heights h with respect to the turbulence developed in the whole troposphere (Table \ref{J}-left side) as well as the percentage of turbulence developed in the (0, $h$) range with respect to the turbulence developed in the whole troposphere (Table \ref{J}-right side). Table \ref{Jd} shows the same calculation obtained in case the median dome seeing is subtracted.

\begin{table}
{\begin{tabular}{@{}cccc@{}}
\hline
Bins &'Good'  &'Typical' &'Bad'  \\
(m) &  J (m$^{1/3}$) & J (m$^{1/3}$) & J (m$^{1/3}$) \\
\hline
14000-20000 & 7.61e-15 &  1.02e-14 &  1.91e-14 \\
12000-14000 & 5.10e-15 &  9.65e-15 &  1.57e-14 \\
10000-12000 & 7.24e-15 &  1.27e-14 &  2.89e-14 \\
8000-10000 &  8.52e-15 &  1.72e-14 &  3.22e-14 \\
6000-8000 & 6.45e-15 &  1.05e-14 &  1.77e-14 \\
4000-6000 &  9.50e-15 &  1.23e-14 &  2.10e-14 \\
3000-4000 &  9.18e-15 &  1.14e-14 &  1.45e-14 \\
2000-3000 &  1.97e-14 &  3.39e-14 &  4.02e-14 \\
1500-2000 & 6.98e-15 &  1.52e-14 &  2.86e-14 \\
1000-1500 &  5.28e-15 &  1.50e-14 &  2.86e-14 \\
\hline
\end{tabular}}
\caption{Composite profiles for h $>$ 1 km. In the first column the boundaries of the vertical slabs. In each column is reported the value of J that is proportional to the integral of the optical turbulence in the correspondent vertical slab. These composite profiles are statistically representative for 43 nights.}
\label{comp_h_gt_1km}
\end{table}

\begin{table}
{\begin{tabular}{@{}cccc@{}}
\hline
 Ground Layer Seeing& 'Good' &'Typical' & 'Bad'  \\
\hline
GS  & 0.55 & 0.81 & 1.16 \\
WB  & 0.62 & 0.75 & 0.88 \\
f$_{gl}$  & 0.82 & 1.1 & 1.6 \\
\hline
\end{tabular}}
\caption{f$_{gl}$ factor for the 'Good', 'Typical' and 'Bad' distribution.}
\label{fgl_factor}
\end{table}

\begin{table}
{\begin{tabular}{@{}cccc@{}}
\hline
 Bins & 'Good'&  'Typical'& 'Bad'  \\
(m) & J (m$^{1/3}$) &  J (m$^{1/3}$) & J (m$^{1/3}$) \\
\hline
900-1000 & 4.35e-15 &  7.06e-15 &  1.33e-14 \\
800-900 & 2.48e-15 &  3.53e-15 &  6.29e-15 \\
700-800 & 3.30e-15 &  6.04e-15 &  1.04e-14 \\
600-700 & 5.72e-15 &  1.09e-14 &  2.05e-14 \\
500-600 & 4.30e-15 & 7.70e-15 &  1.38e-14 \\
400-500 & 4.21e-15 &  9.60e-15 &  1.86e-14 \\
300-400 &  2.05e-14 &  4.26e-14 &  8.26e-14 \\
200-300 & 6.15e-15 &  1.14e-14 &  2.53e-14 \\
100-200 & 2.03e-14 &  4.14e-14 & 8.11e-14 \\
0-100 & 1.79e-13 &  3.27e-13 &  6.17e-13 \\
\hline
0-100 &  3.4e-14 &  8.5e-14 &  2.20e-13 \\
\hline
\end{tabular}}
\caption{Composite profiles for h $<$ 1 km after the {\it 'normalization'} for the f$_{gl}$ factor. These composite profiles are statistically representative for 43 nights. In the last line are reported the J values obtained without the dome contribution (median values: $\varepsilon_{d,25}$ = 0.35 arcsec, $\varepsilon_{d,50}$ = 0.52 arcsec, $\varepsilon_{d,75}$ = 0.70 arcsec).}
\label{comp_h_lt_1km_corr}
\end{table}

\begin{table}
{\begin{tabular}{@{}cccccc@{}}
\hline
 h & J$_{[h,top]}$/J$_{tot}$ &&  h & J$_{[0,h]}$/J$_{tot}$\\
 (m)  &  (\%)  &&  (m) & (\%)\\
\hline
1000  & 25  && 1000 & 75 \\
800  & 26 &&   800&  74\\
600  & 29  && 600 & 71\\
400  & 31 &&  400& 69\\
300  & 38  && 300 & 62\\
200  & 40  &&  200& 60\\
100  & 47 &&  100 & 53\\
\hline
\end{tabular}}
\caption{{\bf Dome seeing included -} {\bf Left:} Percentage of turbulence developed above the height $h$ with respect to the turbulence developed on the whole troposphere. {\bf Right:} Percentage of turbulence developed between the ground and the height h with respect to the turbulence developed in the whole turbulence. Second and fourth columns are obviously complementary.}
\label{J}
\end{table}

\begin{table}
{\begin{tabular}{@{}cccccc@{}}
\hline
 h & J$_{[h,top]}$/J$_{tot}$ &&  h & J$_{[0,h]}$/J$_{tot}$\\
 (m)  &  (\%)  &&  (m) & (\%)\\
\hline
1000  & 40  && 1000 & 60 \\
800  & 43 &&   800&  57\\
600  & 47  && 600 & 53\\
400  &52 &&  400& 48\\
300  & 63  && 300 & 37\\
200  & 66  &&  200& 34\\
100  & 77 &&  100 & 23\\
\hline
\end{tabular}}
\caption{{\bf Dome seeing subtracted -} {\bf Left:} Percentage of turbulence developed above the height $h$ with respect to the turbulence developed on the whole troposphere. {\bf Right:} Percentage of turbulence developed between the ground and the height h with respect to the turbulence developed in the whole turbulence. Second and fourth columns are obviously complementary.}
\label{Jd}
\end{table}

\begin{figure}
\centering
\includegraphics[width=7.5cm]{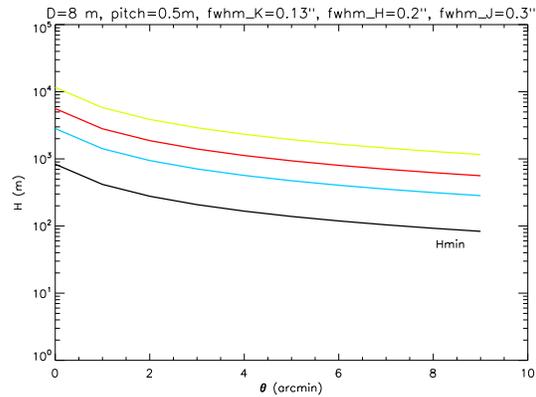}
\caption{Extent of the {\it 'gray zone'} i.e. H$_{min}$ $<$ h $<$ H$_{max}$ for different fields of view and wavelengths in the case of median distribution (50$\%$). H$_{max}$ (colored lines) is calculated for J (blue), H (red) and K (yellow) band. H$_{min}$ (black line) is the same for all the wavelengths. The pupil size is D = 8 m and the pitch size $\Delta$X = 0.5 m. The FWHM is equivalent to r i.e. the residual wavefront coherence size after correction.
\label{glao}} 
\end{figure}

\begin{table*}
{\begin{tabular}{@{}cccccccccc@{}}
\hline
75$\%$  & FWHM& H$_{max}$ &50$\%$  & FWHM& H$_{max}$ &  25$\%$ & FWHM & H$_{max}$ \\
  & (arcsec) & (m) &  & (arcsec) & (m) &  & (arcsec)& (m) \\
\hline
J & 0.43& 378 &J & 0.30& 567 & J  & 0.18 & 945 \\
H & 0.37&606&H & 0.20&1123 &  H& 0.11&2042 \\
K & 0.25&1208 &K & 0.13&2324 & K & 0.08  &3777 \\
\hline
\end{tabular}}
\caption{Values of H$_{max}$ calculated for $\theta$ = 4 arcmin and different residual FWHM. The FWHM values are obtained with the GLAO simulations using as inputs Table \ref{comp_h_gt_1km} and Table \ref{comp_h_lt_1km_corr}. 
H$_{min}$ = $\Delta$X/2$\theta$ $\sim$ 200 m for all the wavelengths.
\label{glao_tab}}
\end{table*}

The composite distribution of the $\CN2$ permits also to quantify the extent of the {\it gray zone}. We note, firstly, that H$_{max}$ = r/$\theta$ (where r is the residual wavefront coherence size after GLAO correction i.e. the residual FWHM) depends on the wavelength.  Using the $\CN2$ composite profiles extended on 20 km as an input (Table \ref{comp_h_gt_1km} and Table \ref{comp_h_lt_1km_corr}), with the combination 'good'-'good', 'typical'-'typical' and 'bad'-'bad', in the boundary layer and in the free atmosphere, the ARGOS simulations provide the residual values r (L. Busoni, private communication). From these values of r that we reported in Table \ref{glao_tab} we can retrieve the values of H$_{max}$ for the wavelengths in the near-infrared range: J, H and K (Table \ref{glao_tab}). Knowing that H$_{min}$ = 200 m, as explained previously, we deduce that the {\it 'gray zone'} extends in the (200 m - 378 m) range and assume its smallest value when we observe in J band and we consider the 'bad'-'bad' case (75\% case). It extends in the (200 m - 3777 m) range and assumes its largest value when we observe in K band and we consider the 'good'-'good' case (25\% case). Fig. \ref{glao} shows the H$_{max}$ (blue, red and yellow lines) for different field of view in the case the residual FWHM of GLAO simulations has been obtained with the central column of Table \ref{comp_h_gt_1km} and Table \ref{comp_h_lt_1km_corr} i.e (case 50$\%$).

\section{Optical turbulence vertical distribution: $\CN2$}
\label{cn2}

\subsection{GS: vertical distribution on the whole troposphere}

\begin{figure*}
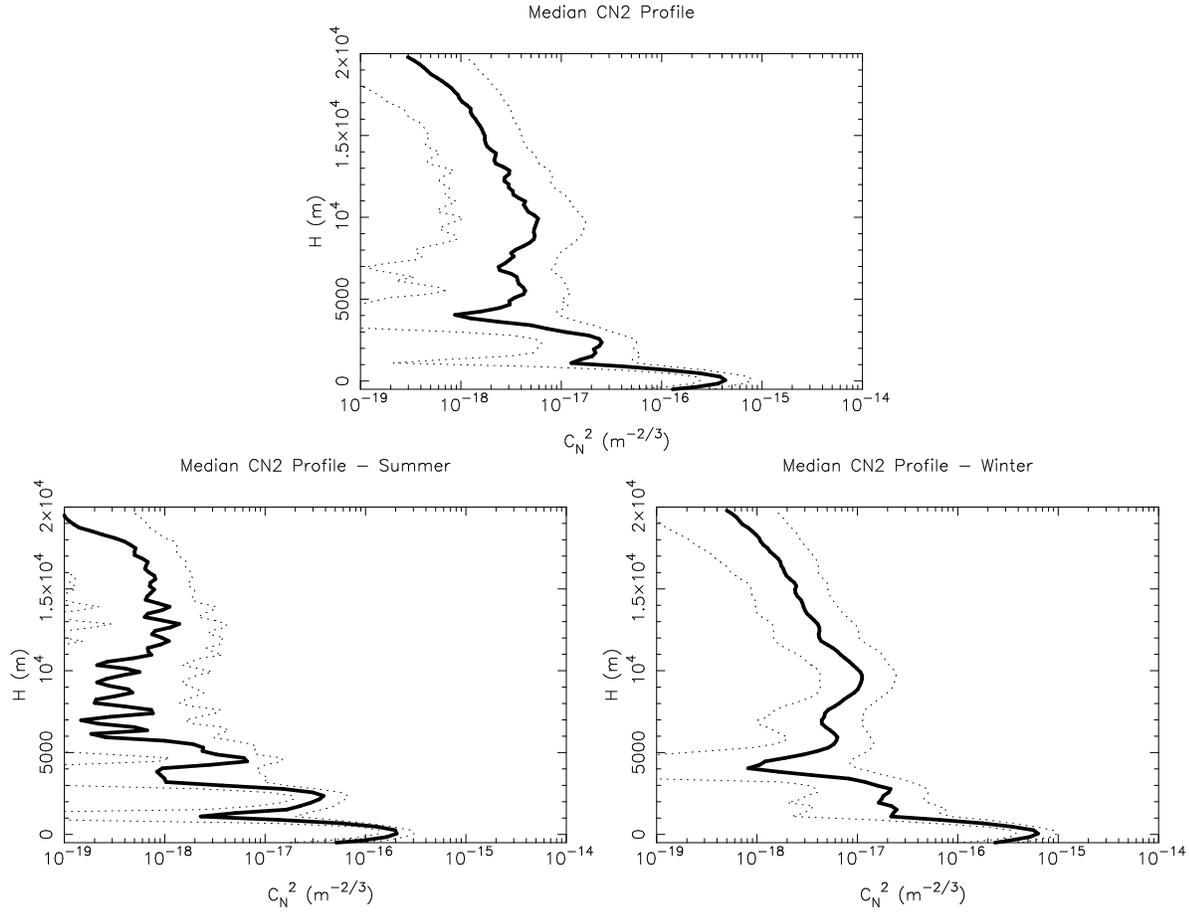

\centering
\includegraphics[width=6cm,angle=-90]{masciadri_fig4a}

\includegraphics[width=6cm,angle=-90]{masciadri_fig4b}
\includegraphics[width=6cm,angle=-90]{masciadri_fig4c}
\caption{Median $\CN2$ profile obtained with the complete sample of 43 nights, the summer [April-June] and winter [October-March] time samples. Results are obtained with the standard GS technique.
\label{cn2_median}} 
\end{figure*}

Figure \ref{cn2_median} shows the median $\CN2$ profile obtained with the whole data-set of 43 nights, the summer and the winter periods. The morphology of the vertical distribution of the optical turbulence ($\CN2$ profile) shows that the greatest turbulence contribution develops in the first kilometer above the ground. Between 1 and 10 km we observe a set of minor peaks changing their position and strength during the year. At around 10 km we observe the typical secondary $\CN2$ peak developed at the jet-stream level. 
For what concerns the seasonal variation we observe that the ground layer bump, responsible for most of the turbulence budget, shows a clear seasonal trend indicating larger turbulence strength in winter than in the summer period.  In the free atmosphere we observe the interesting effect of the secondary $\CN2$ peak located at 10 km in winter time that shifts to higher heights ($\sim$ 14 km) and is characterized by a weaker strength in summer time. The latter effect (that we call {\it '$\alpha$ effect'}) has been put in evidence the first time by (Masciadri \& Egner 2006) above a different site (San Pedro M\'artir) with simulations provided by a mesoscale atmospherical model. Above Mt. Graham the jet-stream $\CN2$ peak appears to have a similar shift of $\sim$ 4 km toward higher heights in summer and the $\CN2$ peak is located at roughly the same absolute height from the ground with respect to San Pedro M\'artir (Masciadri \& Egner, 2006). At that time we had no measurements extended on different periods of the years above San Pedro M\'artir to retrieve a seasonal trend. Avila et al. (2004) referred just to the spring and therefore they could not put in evidence any seasonal trend. 
It is worth to note that this was, to our knowledge, the first time that an atmospherical model could put in evidence new insights before measurements showing that simulations can be a valuable tool to investigate the nature of the turbulence by itself. 

More recently (Els et al. 2009) MASS measurements done above San Pedro M\'artir and extended on yearly time scale have been published as part of the TMT site testing project (Schoeck et al. 2009). The MASS vertical resolution is much lower than the GS one but some information, useful in this context, can be retrieved. The layer at 8 km (the nearest to the jet-stream level) is unfortunately not perfectly centred on the jet-stream and considering a resolution of 4 km at this height the absolute values of the peak-to-peak seasonal variation can be smoothed out. However, if we do not take care about the absolute estimate of the amplitude peak-to-peak variation for which the MASS is not the most suitable instrument, we can say, in any case, that the MASS 8 km layer can provide a qualitative seasonal trend near the jet-stream level. In Els et al. (2009) - Fig.5 we can observe above San Pedro M\'artir a seasonal effect of the turbulence strength similar to what has been obtained above the same site with simulations performed with atmospheric models (Masciadri \& Egner 2006) and to what has been observed above Mt. Graham i.e. an increase of the turbulence strength in the local winter at the jet-stream level. A similar trend is evident in Els et al. (2009)-Fig.5 also for all the other astronomical sites. It is worth to note that in the sites in the south hemisphere, the seasonal trend of the 8 km layer is inverted i.e. we found the maximum values in proximity of the local winter. This effect seems therefore confirmed so far above all the astronomical sites for which measurements are available. 
However, the MASS is not the suitable instrument to put in evidence the more complex effect ('$\alpha$ effect') of the vertical shift of the jet-stream $\CN2$ peak towards higher heights in summer as is resolved by the GS. Also the lack of seasonal trend of the layer located at 16 km in the Els et al. (2009) paper does not mean that there is no seasonal variations at this height but simply that this instrument can not resolve seasonal variations  since its vertical resolution is $\sim$ 8 km at h = 16 km. We conclude therefore that results obtained at San Pedro M\'artir and at Mt. Graham with a GS are an evidence that the {\it '$\alpha$ effect'} exists and it is not typical of a specific site.

Similar behaviors of the $\CN2$ observed above different sites is promising for describing a {\it 'universal physical model'} able to explain what is the origin of the $\alpha$ effect. Masciadri \& Egner (2006) proposed an explanation.

\begin{figure*}
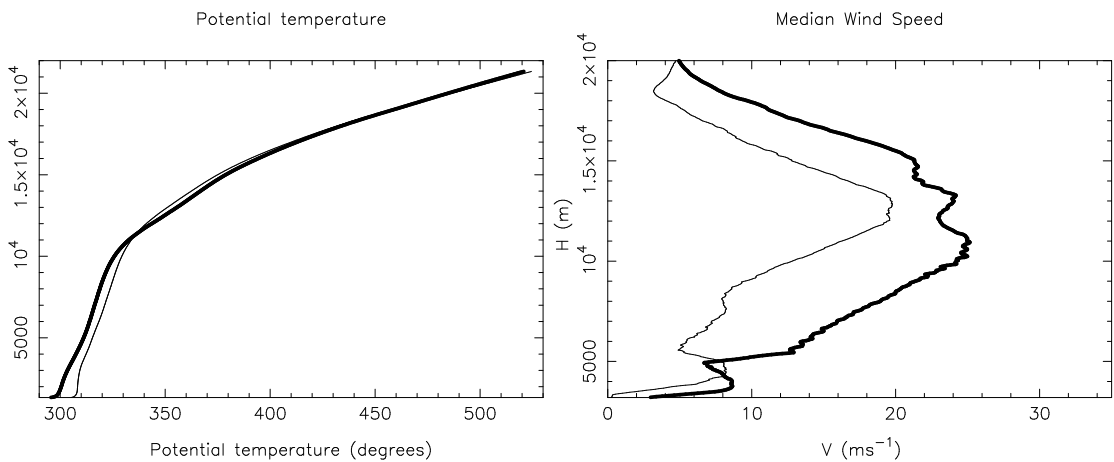

\includegraphics[width=6cm,angle=-90]{masciadri_fig5a}
\includegraphics[width=6cm,angle=-90]{masciadri_fig5b}
\caption{{\bf Left:} Median potential temperature calculated using the ECMWF analyses. {\bf Right:} Median wind speed calculated using the composite profile (GS below 2km, ECMWF analyses above 2 km - see text) in summer (thin style line) and winter (bold style line).
\label{TV_graham}}
\end{figure*}

Starting from the assumption that the development of the optical turbulence depends on the gradient of the wind speed and the potential temperature a parallel analysis of these two elements all along the troposphere in the two seasons can provide insights on the seasonal variations trends.  
Above San Pedro M\'artir no great difference have been identified in the potential temperature and its gradient between the summer and winter. On the other hand a substantial decrease of the gradient of the wind speed and its strength has been identified at a jet-stream level in summer. This fact could explain the decrease of the turbulence strength in summer at a jet-stream level. Besides, the rapid inversion of the wind speed gradient at around 15-16 km could explain the production of the optical turbulence at these heights that appears as a shift of the jet-stream $\CN2$ peak towards higher heights.

Figure \ref{TV_graham} shows the median potential temperature and wind speed profiles above Mt. Graham calculated in the two periods (winter and summer) in which $\CN2$ measurements have been collected. The potential temperature is retrieved from the ECMWF analyses in the  same grid point we discussed in Section \ref{astro_integ}. The wind speed profiles are obtained with the composite procedure described in Section \ref{astro_integ} i.e. GS measurements have been considered for h $<$ 2 km and ECMWF analyses for h $>$ 2 km. We observe a not negligible wind speed increase at the jet-stream level in winter time as already observed above San Pedro M\'artir. Also a similar inversion in the wind speed gradient in the high part of the atmosphere above the jet-stream (at $\sim$ 17 km a.s.l.) is visible above Mt. Graham as well as above San Pedro M\'artir in summer. Above Mt. Graham the median profile of the potential temperature shows that the typical {\it 'slope change'} identifying the position of the tropopause visible at 10 -11 km in winter time, is a little smoother and located at higher heights ($\sim$ 13 km a.s.l.) in summer. The level of the thermal stability plays obviously an important role to determine the strength of the optical turbulence triggered by the dynamic shear typical of the jet-stream. This fact might therefore play a role in the mechanism producing the shift towards higher heights of the jet-stream $\CN2$ peak but is, in any case, coherent with what found above San Pedro M\'artir.  The physical model proposed by Masciadri \& Egner (2006) to explain the $\alpha$ effect seems therefore to be consistent with what is observed above Mt. Graham in this paper. We note that, the fact that the wind speed is visibly one of the main causes triggering the seasonal variation of the $\CN2$ at the jet-stream level does not mean that for equivalent wind speeds above whatever astronomical site, one has to expect the same $\CN2$ value. The absolute strength of the $\CN2$ depends in a precise region of the atmosphere, indeed, on the thermodynamic stability of the atmosphere in the same region.

\subsection{HVR-GS: vertical distribution for h $\le$ 1 km}

The optical turbulence  vertical distribution with high resolution (20-30 m) in the first kilometer is obtained with the method called HVR-GS presented in Egner \& Masciadri (2007). The HVR-GR data-set is obtained taking the integral of the $\CN2$ profiles retrieved from the AC frames and redistributing the energy in the first kilometer according to the detected triplets in the CC frames.
Three different strategies can be used to study the turbulence spatial distribution in the boundary layer. The usefulness of each method depends on the application one intends to give to the analysis. We study:\newline\newline
(A)  the median of the $\CN2$ (and/or J) profiles.\newline
(B)  the average of the $\CN2$ (and/or J) profiles. This method is very useful for comparisons of measurements and simulations obtained with atmospherical models. Moreover this operator has the advantage that the mean of J$_{i}$ is equal to the J retrieved from the mean of the C$_{N,i}^{2}$ (where J$_{i}$ and C$_{N,i}^{2}$ refer to the each individual profile). It is not the case for the median.  \newline
(C) the composite profiles as calculated in Section \ref{composit}. This method is very useful for applications to Adaptive Optics.\newline

\begin{figure*}
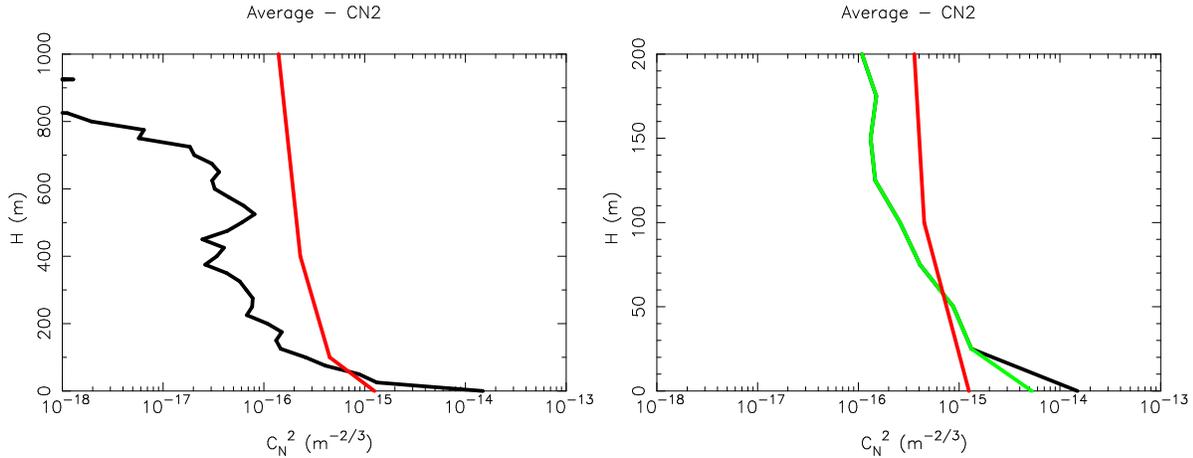

\centering
\includegraphics[width=6cm,angle=-90]{masciadri_fig6a}
\includegraphics[width=6cm,angle=-90]{masciadri_fig6b}
\caption{{\bf Left:} Mean $\CN2$ profile ($\Delta h$$\sim$20-30 m) calculated with the sample 'HVR-GS' and the dome seeing contribution included (black line). Hufnagel model (red line). {\bf Right:} Zoom of the first 200 m. Near the ground it is put in evidence  the different $\CN2$ shape in case the dome contribution is included (black line) and excluded (green line). 
\label{hvr-gs}} 
\end{figure*}

Following the strategy (A) we find that $\CN2$ = 0 for h $>$ 25-30 m. This can be explained with the fact that each vector ($\CN2$ profile) has many zeros. This is due to the fact that, during the monitoring, the system detects spikes and thin layers that changes position and duration with the time. In other words, the use of the median in these cases, can be misleading and it can provide very low values of the $\CN2$. Considering that the strategy (A) is not really useful to characterize these measurements in our case, the strategies (B) and (C) have been used. 
Figure \ref{hvr-gs}-left shows the result obtained following the strategy (B) i.e. the mean of the $\CN2$ profiles ($\Delta$h $\sim$ 25-30 m) calculated from the sample 'HVR-GS' after normalization for the f$_{gl}$ factor and it is therefore representative of 43 nights. The two profiles (with and without dome contribution) are shown in proximity of the ground (Figure \ref{hvr-gs}-right). To retrieve the typical scale height B of the exponential decay of the mean $\CN2$ profile the measurements done below 125 m have been fitted with an exponential law (Eq.(\ref{eqn:exp})) as we already did in Stoesz et al. (2008):

\begin{equation}
y=A \cdot e^{(-h/B)}
\label{eqn:exp}
\end{equation}

where A and B are free parameters. The calculation is obviously shown only in the case in which the dome contribution is subtracted. The fit gives A $=$ 3.34$\cdot$10$^{-15}$ and B $=$ 37.4 m (Fig.\ref{hvr-gs}-left). If we limit the analytical fit to the first 30 meters, the scale height B = 28 m. The exponential decay is however an absolutely arbitrary analytical law. The important issue if that these results definitely indicate that the HVR-GS technique is able to put in evidence that the turbulence decays above typical astronomical sites in stable night time conditions in a much sharper way than what has been predicted and quantified in the past. Indeed, the Hufnagel model in proximity of the surface (Roddier, 1981) in night conditions states that the turbulence scales as h$^{-2/3}$ (Fig.\ref{hvr-gs}). For the Hufnagel model (Fig.\ref{hvr-gs}-left)  the $\CN2$ decreases of one order of magnitude within 1 km while our results indicate that the $\CN2$ decreases of one order of magnitude within $\sim$ 60-70 m (Fig.\ref{hvr-gs}-right). It remains interesting the (800 m - 1 km) range in which visibly a weak turbulence develops. One should expect a smoother connection between the boundary layer and the free atmosphere. We think that this is just an artifact effect due to the use of a quite different resolution below and above 1 km. From the point of view of the AO simulations this small gap should not cause any problems. It should be enough to implement a weak convolution to slightly smooth out the $\CN2$ vertical profiles at the interface located at 1 km so to obtain a less abrupt connection of the $\CN2$ profile above and below 1 km. 

We note that the presence of no signal (therefore $\CN2$= 0) might potentially indicate that the turbulence strength is not equal to zero but simply weaker than the threshold $\CN2$$=$10$^{-16}$ (associated to an equivalent J$=$2.5$\cdot$10$^{-15}$).  
In Masciadri et al. (2009) it has been calculated, in a post-processing phase, the most conservative case in which we assigned $\CN2$$=$10$^{-16}$ where there is no signal. After a more careful investigation we observed that the lattest distribution is associated to a too large total J in the boundary layer and, for this reason, it can be discarded. 

\begin{figure*}
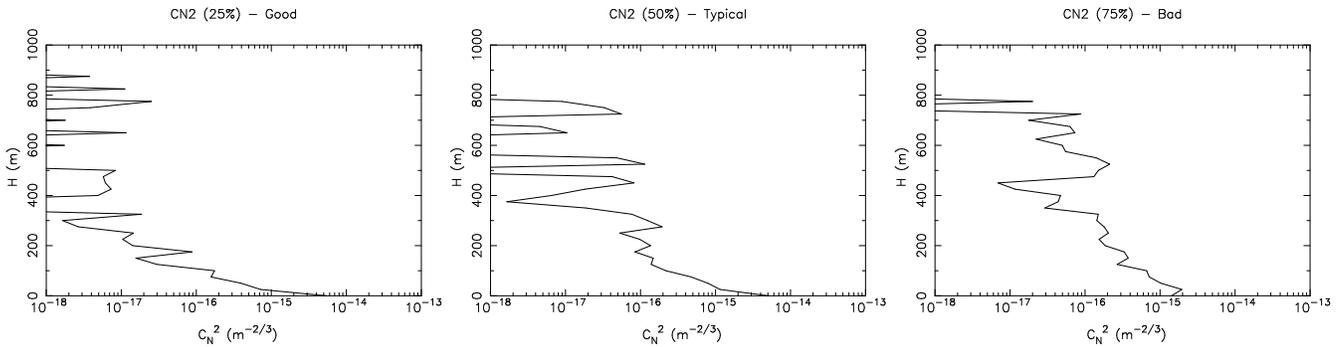

\centering
\includegraphics[width=4.5cm,angle=-90]{masciadri_fig7a}
\includegraphics[width=4.5cm,angle=-90]{masciadri_fig7b}
\includegraphics[width=4.5cm,angle=-90]{masciadri_fig7c}
\caption{Mean $\CN2$ profiles calculated from the corresponding J (or r$_{0}$) values related to the 20-30 $\%$, 45-55 $\%$ and 70-80 $\%$ ranges of the J cumulative distribution. The dome contribution is excluded. In Annex C the numerical values for the correspondent J values are reported.} 
\label{cn2_25_50_75}
\end{figure*}

Figure \ref{cn2_25_50_75} shows the results obtained following the strategy (C) that is the mean $\CN2$ profiles associated to the J (or r$_{0}$) related to the 20-30 $\%$, 45-55 $\%$ and 70-80 $\%$ of the cumulative distribution. In Annex C the numerical values for the correspondent J values are reported. Curiously the first grid point near the ground of the 75$\%$ case distribution shows a weaker value with respect to the 25$\%$ and 50$\%$ cases. This is due to the fact that the third quartile dome seeing ($\varepsilon_{d,75}$), that has been subtracted from the original value is particularly large (0.70 arcsec). The morphology of the $\CN2$ in the first kilometer is very interesting showing several thin layers and a very weak turbulence between 800 m and 1 km. We note that the sample on which we calculate the average in each slot [20-30]$\%$, [45-55]$\%$ and [70-80]$\%$ is of a few hundreds of $\CN2$ profiles therefore there are no doubts that this structure reproduces some real distrubution. In terms of morphology of the turbulence profile we find therefore that the higher the vertical resolution the thinner the size of the detected layers. This conclusion is perfectly coherent with the turbulence structure resolved by balloons equipped for the $\CN2$ measurements (Azouit \& Vernin, 2005). 

\begin{figure*}
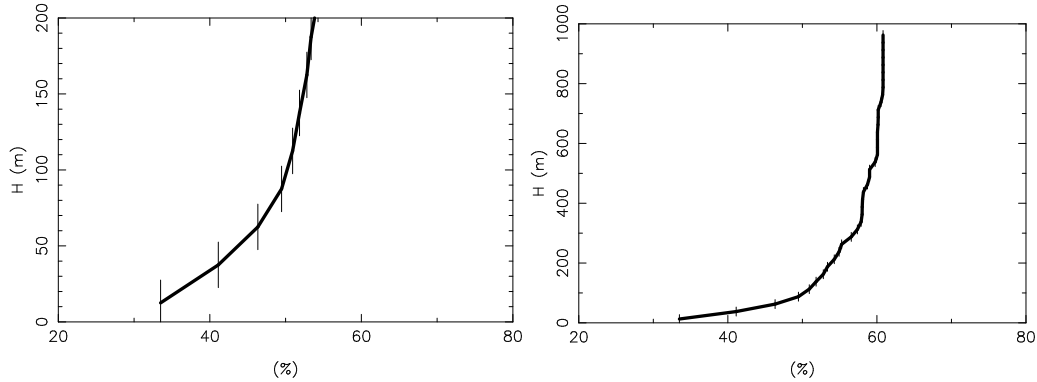

\centering
\includegraphics[width=5cm,angle=-90]{masciadri_fig8a}
\includegraphics[width=5cm,angle=-90]{masciadri_fig8b}
\caption{Percentage of turbulence developed between the ground and the height $h$ with respect to the turbulence developed in the whole atmosphere ($\sim$ 20 km) as retrieved from the HVR-GS measurements and extended to the first kilometer. On the left hand is shown a zoom of the picture centred on the first hundreds of meters.} 
\label{perc}
\end{figure*}

Figure \ref{perc} shows that the percentage of turbulence P(h) developed in the (0, h) range with h in the (0, 1 km) vertical slab. The function P(h) is defined as: 
\begin{equation}
P(h) = \left( {\frac{{\int\limits_0^h {C_N^2 (h^* )dh^* } }}
{{\int\limits_0^{\infty} {C_N^2 (h^* )dh^* } }}} \right) \times 100
\label{perc_eq}
\end{equation}

where the $\CN2$ profile is that associated to the 45-55 $\%$ case (Table \ref{comp_h_lt_1km_hvr} and Fig. \ref{cn2_25_50_75}-centre). It is worth to note that the error bars for the HVR-GS technique is of the same order of half of the vertical resolution i.e. $\pm$ 12-15 m. This derives mainly by the definition of the zero point i.e. the ground that is characterized by an uncertaintly equivalent to half of the vertical resolution. 

From Fig.\ref{perc} we retrieve that around 50$\%$ of the turbulence developed in the whole 20 km is concentrated below 80$\pm$15m.  This result is substantially different from the preliminary indications obtained in Egner \& Masciadri (2007) and the turbulence seems to be much more concentrated near the ground. We also note that the $\CN2$ morphology below 1 km decreases in a different way if we look at the sample 'WB' and the 'HVR-GS'. This is not a contradiction and it can be explained with the fact that the turbulence spatial distribution for the 'WB' and the 'HVR-GS' samples is not necessarily the same because the first is based on the AC frames while the second is based on the CC frames. It is interesting to note that the higher the vertical resolution, the sharper is the exponential decay of the morphologic turbulence structure. 

\begin{figure}
\centering
\includegraphics[width=6cm,angle=-90]{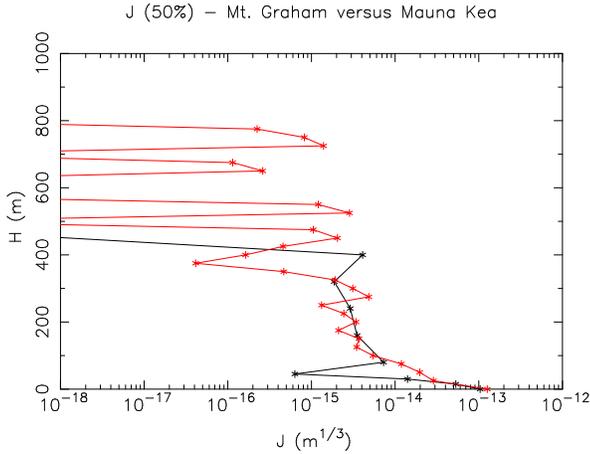}
\caption{Red line: turbulence vertical distribution (J-profile, 50$\%$ case) above Mt. Graham as calculated in Fig.\ref{cn2_25_50_75}-centre. Black line: turbulence vertical distribution (J-profile, 50$\%$ case) above Mauna Kea as calculated by Chun et al. (2009) in Table 6-centre. Mauna Kea measurements extend up to 650 m.} 
\label{mg_vs_mk}
\end{figure}

We note that, simultaneously to our study, some other authors (Chun et al. (2009)) recently investigated the turbulence structure near the ground at high vertical resolution. Even if the instruments employed were different these studies present many similarities in the results. From a qualitative point of view, we note that, also in that case, the turbulence appears well confined near the surface. From a quantitative point of view, things are more delicate. Looking at Table 3 (in that paper) it appears that their data-reduction is more similar to our method (C) than to the others methods. Figure \ref{mg_vs_mk} shows the J-profile retrieved from Fig. \ref{mg_vs_mk} - centre overlapped to the J-profile retrieved from Table 3 (Chun et al. (2009). While above Mt. Graham the turbulence vertical sampling is 25 m, above Mauna Kea  the sampling increases from 15 m up to 80 m and it extends only up to 650 m. The turbulence vertical distribution appears very similar. we note that above Mauna Kea a local minimum is present at $\sim$ 45 m above the ground more or less in correspondence of the abrupt detection break due to the sensitivity threshold from LOLAS (Table 3 - Chun et al. (2009)).  We highlight that the evident  huge {\it 'turbulent vacuum zone'}  between 560 m and 1 km (SLODAR) simply means that the turbulence is not measured in this vertical slab, not that turbulence is not present.

\section{Conclusions}
\label{concl}
In this paper we present the results of a study aiming to characterize the optical turbulence at Mt. Graham. We present a general overview of the statistics (43 nights) of the $\CN2$ profiles and all the main integrated astroclimatic parameters and their seasonal trends. The main conclusions we achieved are: \newline
{\bf (1)} With a median seeing $\varepsilon$$=$ 0.95 arcsec ($\varepsilon$$=$ 0.72 arcsec without dome contribution), isoplanatic angle $\theta_0$$=$ 2.5 arcsec and a wavefront coherence time $\tau_0$$=$4.8 msec, Mt. Graham confirms its good quality in terms of turbulence characteristics typical of the best astronomical sites in the world. All the integrated astroclimatic parameters (the seeing, the isoplanatic angle, the wavefront coherence time and the equivalent wind speed) show a clear seasonal trend that indicates better turbulence conditions and weaker equivalent wind speed V$_{0}$ in summer with respect to the winter. \newline
{\bf (2)} The ground layer is characterized for the first time with a high resolution (200-250 m and 20-30 m). The turbulence exponentially decays above Mt. Graham with a much sharper profile than what has been supposed so far and expressed with the Hufnagel model. Three different strategies of analysis aiming to investigate the morphology of the turbulence spatial distribution have been presented. We find that around 50$\% $ of the turbulence developed in the whole atmosphere is concentrated below 80$\pm$15 m from the ground and 60$\%$ of the turbulence in the first kilometer. This evidence together with the favorable large $\theta_{0}$ observed above Mt. Graham (particularly in the spring/summer time) represent extremely favorable conditions for astronomical observations assisted by a LGS/GLAO system such as ARGOS. \newline
{\bf (3)} We observe that the higher is the vertical resolution of the tool used to measured the turbulence vertical distribution the sharper is the turbulence decreasing. \newline
{\bf (4)} The percentage of turbulence developed below the primary mirror of the LBT (i.e. $\sim$ 35 m from the ground) is around 33$\%$. However this estimate has to be considered with precaution because the uncertainty (2$\sigma$$\sim$ 25-30 m) is of the same order of magnitude of the vertical resolution ($\Delta$H $\sim$ 25-30 m) and in the first hundred meters the turbulence decreases very sharply.\newline
{\bf (5)} It appears evident that at Mt. Graham the turbulence decreases above the ground similarly to what observed above Mauna Kea in a more or less simultaneous study (Chun et al. 2009) performed with different instrumentation. \newline
{\bf (6)} A composite distribution of the turbulence on the whole 20 km is calculated to be used as input of AO simulations of the LBT Laser Guide Star system named ARGOS and the calculation of the gray zones for the near-infrared J, H and K band is done. The gray zone extends from a minimum of (200m - 378 m) in J band with the 'bad'-'bad' case up to a maximum of (200m - 3777 m) in K band in the case 'good'-'good' case. \newline
{\bf (7)} A clear $\CN2$ seasonal variation trend has been observed in proximity of the ground and  in the jet-stream regions. These measurements confirm the first evidence of the $\CN2$ seasonal trend observed by Masciadri \& Egner (2006) above other astronomical site. The physical model proposed by Masciadri \& Egner (2006) that is able to explain the seasonal effect of the secondary peak of the $\CN2$ called {\it  '$\alpha$ effect'}, is confirmed and refined.\newline
{\bf (8)} For the first time we observed a seasonal trend of the dome seeing. This is certainly a topic that deserves a more careful investigation in the future.  \newline
{\bf (9)} We proved that the error in the normalization of the scintillation that has been recently put in evidence in the principle of the GS technique affects these measurements with an absolutely negligible quantity (0.04"). In other words, the median seeing retrieved from the GS (without correction) overestimates for $\sim$ 0.04" the correct median seeing. From a general point of view, all the  GS measurements obtained with a pupil size D $\ge$ 1.5 m and a binary separation $\theta$ $\le$ 8 arcsec, are affected by this error of less than a few hundredths of arcseconds. 

\appendix

\section{Normalization of scintillation for the standard Generalized Scidar}

It has been highlighted (Johnston et al. 2002) an imprecision in the normalization of the autocovariance of the scintillation maps obtained with a GS. A generalization of this problem, extended to altitudes h $>$ 0, has been presented more recently by Avila \& Cuevas (2009).  Results of these studies say that, to obtain exact results for the $\CN2$ at an height $h$ from the ground, one has to multiply the $\CN2$(h) retrieved from the GS by a factor 1/(1+$\varepsilon(h)$) where $\varepsilon$ is the relative error between the exact and erroneous autocovariance of the scintillation maps obtained with the GS. $\varepsilon$(h) depends on a set of geometrical parameters related to the optical set up and the observed binaries, more precisely the pupil of the telescope D, the height $d$ at which the detection plane is conjugated below the ground, the ratio between the stellar magnitude of the binary, the angular separation $\theta$ and the central obscuration of the pupil  $e$ expressed as a fraction of the primary mirror. The authors introduced an approximated solution for the correction valid for a simple circular telescope pupil and the exact solution valid for a telescope pupil shape formed by a primary and a secondary mirror. However we calculated that, for our sample, the correction of all the $\CN2$ profiles has an absolutely negligible impact on the statistical analysis presented in this paper (in both cases i.e. approximated and exact solution) and the median $\varepsilon$ calculated on the whole 43 nights sample changes respectively for just 0.03" and 0.04". This means that approximated and exact solution are very similar in this case.

To prove that, we first note that in the LBT-GS, the scintillation maps and the corresponding autocovariance calculated in real time are sampled on the same number of pixels square. If the pupil of the telescope takes the whole field on the CCD, under the condition we have just described, it follows that h$_{max}$ = D/(2$\cdot$$\theta$) (and not D/$\theta$ as indicated by Avila \& Cuveas (2009)). It is known, indeed, that a layer at a height $h$ produces a couple of identical scintillation maps separated by a distance $d$=$\theta$$\cdot$$h$ at which are associated the later peaks of the auto-correlation located at a distance $d$ from the centre of the frame. However, the most external lateral autocorrelation peaks visible on the CCD (that correspond to the highest detectable layer) are necessarily located at a distance d=D/2 (i.e. half of the CCD size) from the centre of the frame if the pupil of the telescope covers the whole CCD and if the support of the auto-correlation is not changed. It follows therefore that h$_{max}$ = D/(2$\cdot$$\theta$). The later peaks associated to turbulent layers located at h $>$ D/(2$\cdot$$\theta$) fall outside the CCD. This means that the vertical range monitored by the LBT-GS is basically half than what is indicated by Avila \& Cuevas (2009).
The important condition to be respected is that h$_{max}$ includes the whole atmosphere and this is the case: the smallest h$_{max}$ is 24 km in the case of the widest binary (7.6 arcsec). We are therefore in a correct configuration to scan the whole atmosphere. Besides we also note that, as shown in Masciadri et al. 2002-Fig.12 for example, the statistical noise increases for h $>$ 20 km (at least for the geometry commonly used by the standard GS and D $>$ 1.5 m) and the GS measurements are no more reliable in that region. 
Looking at Fig. 4 and Fig. 5 (Avila \& Cuevas, 2009) it is possible to appreciate that, for D $>$  1.5 m and $e$ = 0.2 (as it is our case) and for $b$ = $\alpha$/(1+ $\alpha$)$^{2}$, $\alpha$ = 10$^{-0.4\Delta m}$, the value of $\varepsilon$(h) in the exact solution (Sec. 3.2 of that paper) and the approximated one (Sec. 3.1 of that paper) is very small in both cases in the (0, h$_{max}$) range. 

\begin{figure*}
\centering
\includegraphics[width=6cm]{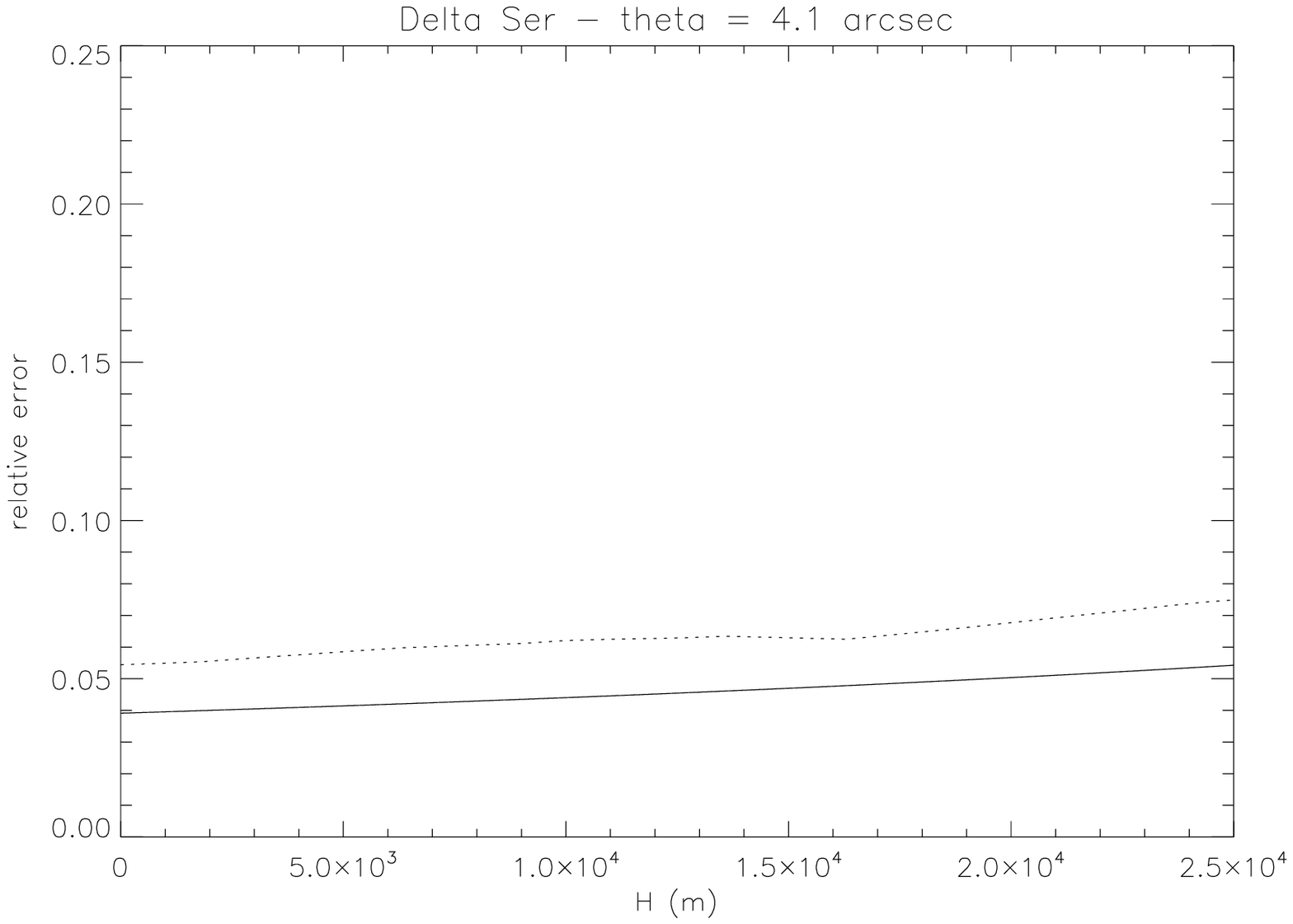}
\includegraphics[width=6cm]{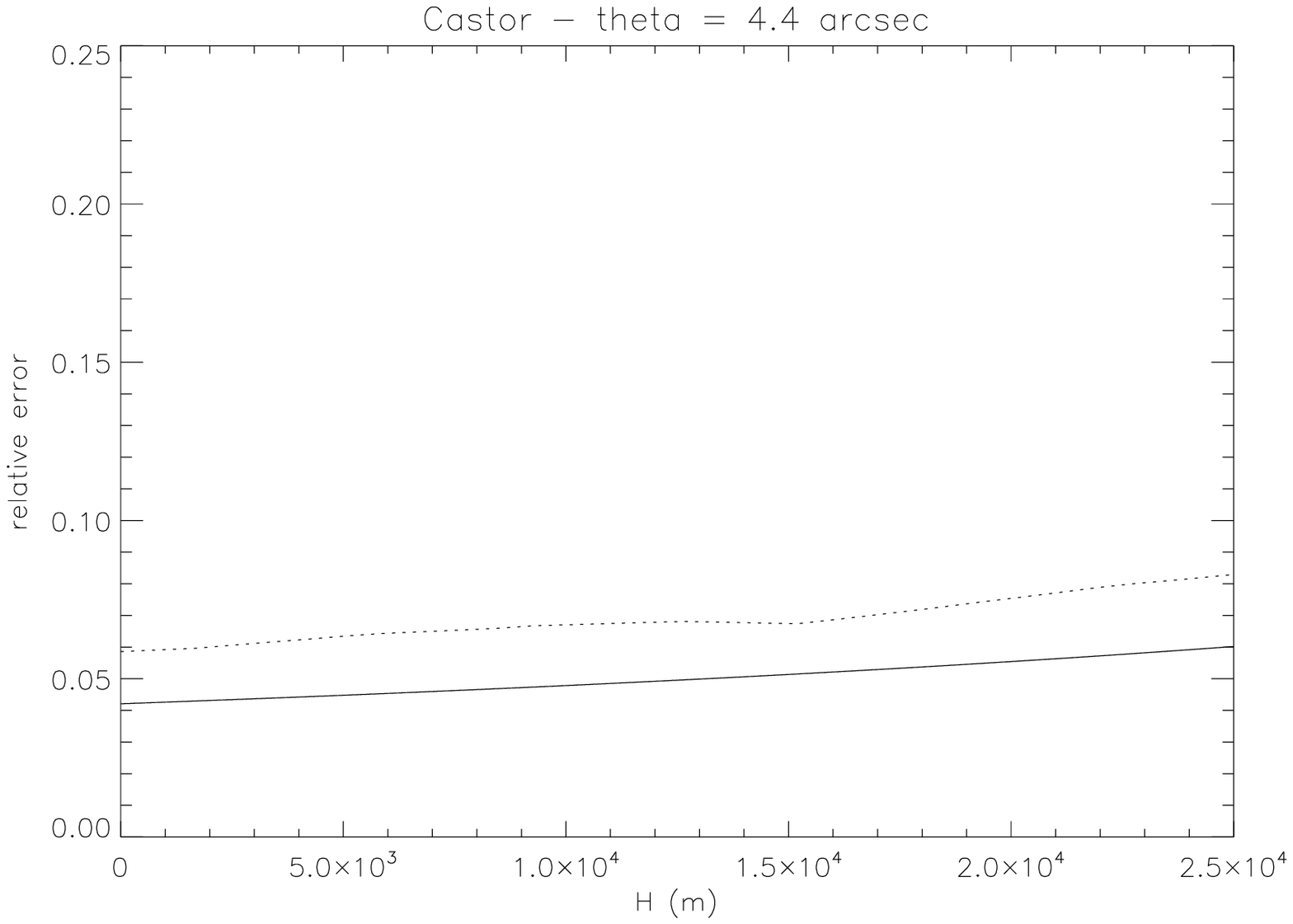}
\includegraphics[width=6cm]{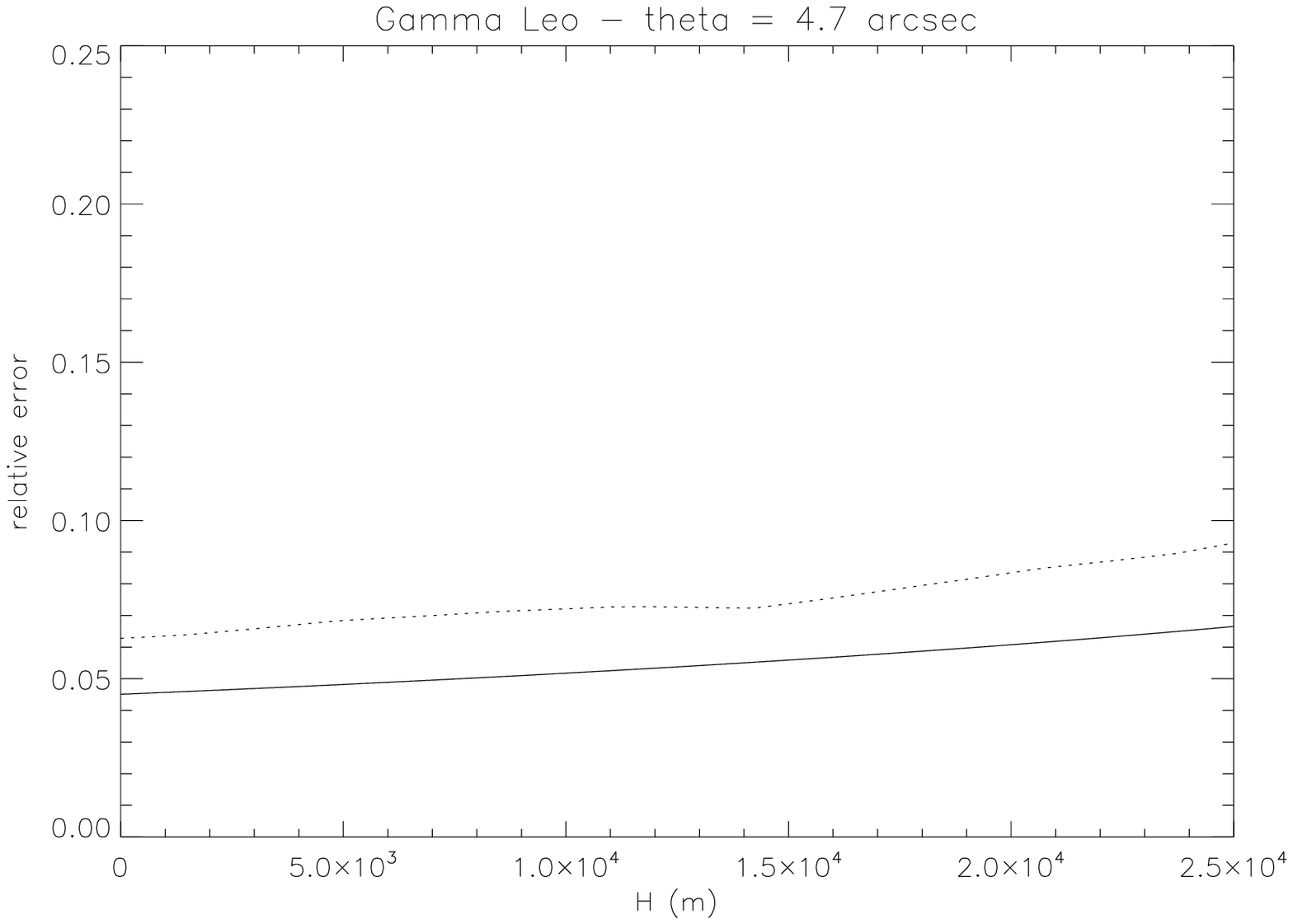}
\includegraphics[width=6cm]{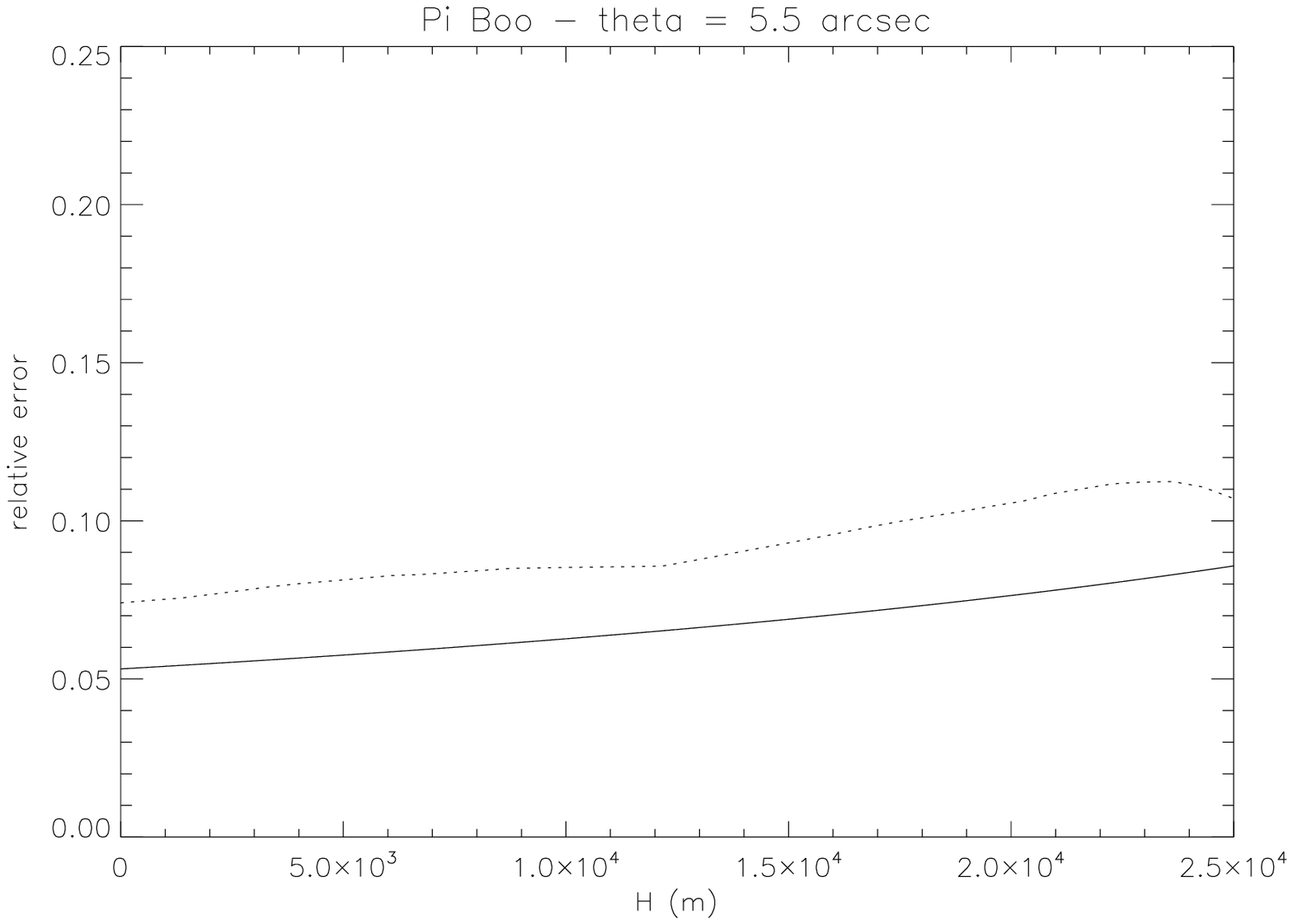}
\includegraphics[width=6cm]{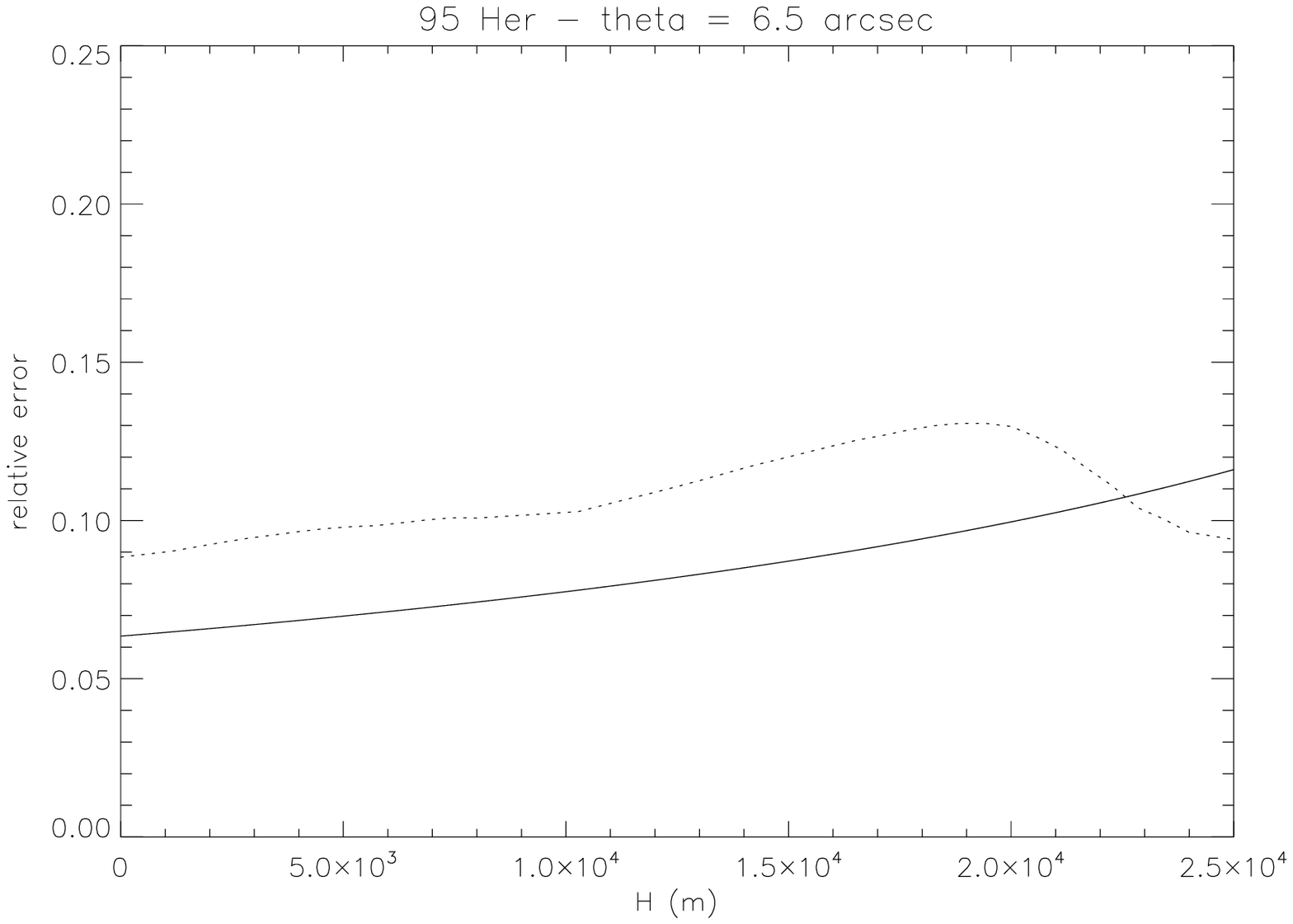}
\includegraphics[width=6cm]{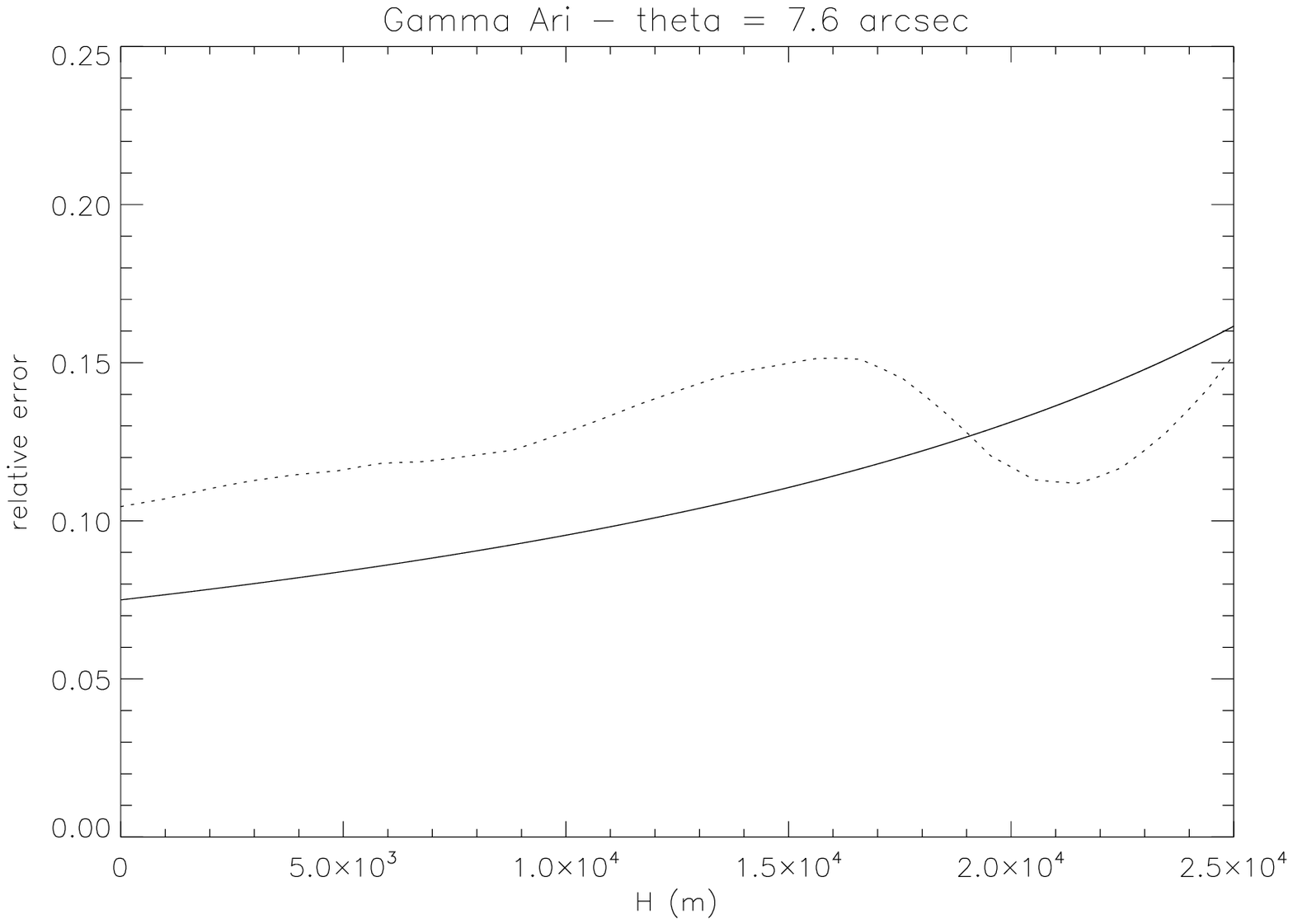}
\caption{Left: The correction factor $\varepsilon(h)$ calculated for D = 1.83 m, d = 3.5 km, $e$=0.2 and the separation of all the binaries of Table \ref{tab_stars}. $\varepsilon(h)$ is the relative error between the exact and erroneous autocovariance of the scintillation maps obtained with the GS. The full thin line is $\varepsilon(h)$ calculated in the approximation of a full circular pupil D=1.83 m without secondary, the dashed line is $\varepsilon(h)$ in the case of a circular pupil D=1.83 m and a secondary D$_{s}$=0.38 m. We note that the relative error has to be multiplied by 100 i.e. 0.1 is equal to 10$\%$.
\label{corr_norm}} 
\end{figure*}

\begin{figure}
\centering
\includegraphics[width=6cm, angle=-90]{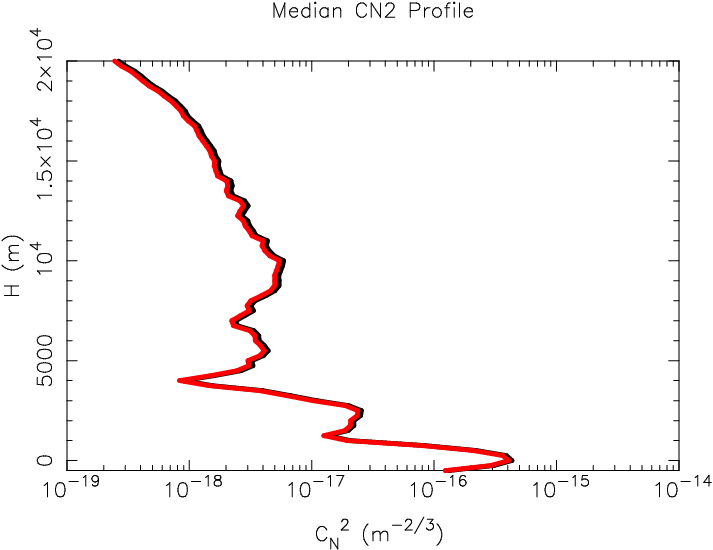}
\caption{Median $\CN2$ profiles related to 43 nights (black line) and corrected by the normalization factor (red line).
\label{corr_cn2}} 
\end{figure}

Figure \ref{corr_norm} shows the relative error $\varepsilon$ as a function of h for all the different binaries used in this study (Table \ref{tab_stars}) and for a fixed value of D = 1.83 m, $e$ = 0.2, d = 3.5 km for the exact solution (full line) and approximated one (dashed one).  We observe that the error $\varepsilon$, in both exact and approximated cases, is typically smaller than 0.1 in the 20 km range and it reaches the value of 0.15 only with the widest binary $\gamma$ Ari ($\theta$ = 7.6"). The error produced on the $\CN2$ reconstructed by the normalization of the autocovariance of the scintillation maps is absolutely negligible in our case as can be seen in Fig.\ref{corr_cn2} producing a difference in the seeing of the order of 0.03" and 0.04" respectively with the approximated and exact solution. 

\section{Normalization of scintillation for the HVR-GS}

The error in the normalization put in evidence by Johnston et al. (2002) and Avila \& Cuevas (2009), in spite of the wider binary employed, affects the HVR-GS measurements in the same negligible way in which the standard GS technique is affected. The reason is that this technique, thanks to the procedure called {\it 'normalization'} (see Sec. \ref{composit}), uses the AC and the CC frames just to retrieve the vertical turbulence distribution in the atmosphere (i.e. the shape). The strength of the turbulence is corrected by the factor f$_{gl}$ that, physically speaking, is equivalent to taking the turbulence measured with the standard GS and redistributing it following the shape of the profile reconstructed by the HVR-GS technique.  The HVR-GS technique retrieves the $\CN2$ profiles in the first kilometer. Let's assume to have an error  A = 1/(1+$\varepsilon$) for the $\CN2$ as indicated by Johnston et al. (2002) and Avila \& Cuevas (2009). We should therefore expect that the corrected turbulence strength is J$_{0}$$^{'}$$_{*}$ = A $\cdot$J$_{0}$$^{'}$ for each vertical grid point of $\sim$ 25 m. However, when we multiply for the f$_{gl}$ factor (Eq.\ref{fgl}), the factor A disappears. Indeed f$_{gl}$ can also be written as J$_{0}$ /(A$\cdot$J$_{0}$$^{'}$). We conclude that, whatever is the value of A, the error potentially introduced by the wide binary is not taken into account in the final value of the $\CN2$ retrieved with the HVR-GS technique. 




\section{Vertical distribution for the HVR-GS: J values}

In Table \ref{comp_h_lt_1km_hvr} are reported the values of J = $\CN2$$\cdot$dh obtained with the average of J included in the 20-30$\%$, 45-55$\%$ and 70-80$\%$ ranges.

\begin{table*}
{\begin{tabular}{@{}cccc@{}}
\hline
H   &'Good'&  'Typical'& 'Bad'  \\
 (m) & J (m$^{1/3}$) & J (m$^{1/3}$) &  J (m$^{1/3}$) \\
\hline
 1000.& 0.000000E+00& 0.000000E+00& 0.000000E+00\\
950.& 0.000000E+00& 0.000000E+00& 0.000000E+00\\
925.&  0.000000E+00& 0.000000E+00& 0.000000E+00\\
900.& 0.000000E+00& 0.000000E+00& 0.000000E+00\\
875.& 0.947308E-16& 0.000000E+00& 0.000000E+00\\
850.& 0.000000E+00& 0.000000E+00& 0.000000E+00\\
825.& 0.279192E-15& 0.000000E+00& 0.000000E+00\\
800.& 0.000000E+00& 0.000000E+00& 0.000000E+00\\
775.& 0.631038E-15& 0.224192E-15& 0.498846E-15\\
750.& 0.974615E-16& 0.822231E-15& 0.000000E+00\\
725.& 0.000000E+00& 0.139250E-14& 0.219769E-14\\
700.& 0.447308E-16& 0.000000E+00& 0.443846E-15\\
675.& 0.000000E+00& 0.114500E-15& 0.157692E-14\\
650.& 0.289577E-15& 0.261962E-15& 0.183635E-14\\
625.& 0.000000E+00& 0.000000E+00& 0.558077E-15\\
600.& 0.435769E-16& 0.000000E+00& 0.124112E-14\\
575.& 0.000000E+00& 0.000000E+00& 0.137462E-14\\
550.& 0.000000E+00& 0.120673E-14& 0.352438E-14\\
525.& 0.000000E+00& 0.284692E-14& 0.532192E-14\\
500.& 0.208654E-15& 0.000000E+00& 0.377000E-14\\
475.& 0.144000E-15& 0.105146E-14& 0.328038E-14\\
450.& 0.153692E-15& 0.204692E-14& 0.172654E-15\\
425.& 0.182692E-15& 0.458077E-15& 0.290846E-15\\
400.& 0.122231E-15& 0.162846E-15& 0.118231E-14\\
375.& 0.000000E+00& 0.413077E-16& 0.109462E-14\\
350.& 0.000000E+00& 0.467115E-15& 0.728231E-15\\
325.& 0.465385E-15& 0.191885E-14& 0.375588E-14\\
300.& 0.413077E-16& 0.312077E-14& 0.359250E-14\\
275.& 0.668462E-16& 0.485431E-14& 0.448385E-14\\
250.& 0.362192E-15& 0.132169E-14& 0.512423E-14\\
225.& 0.261346E-15& 0.243788E-14& 0.384731E-14\\
200.& 0.355769E-15& 0.341577E-14& 0.463577E-14\\
175.& 0.219654E-14& 0.210588E-14& 0.832846E-14\\
150.& 0.391923E-15& 0.369450E-14& 0.943177E-14\\
125.& 0.748615E-15& 0.346750E-14& 0.672115E-14\\
100.& 0.437669E-14& 0.544700E-14& 0.165801E-13\\
 75.& 0.393888E-14& 0.118840E-13& 0.180021E-13\\
 50.& 0.988535E-14& 0.197208E-13& 0.256638E-13\\
 25.& 0.179186E-13& 0.287454E-13& 0.488174E-13\\
 0.& 0.280439E-12& 0.368246E-12& 0.432148E-12\\
\hline
0.& 0.135920E-12& 0.126611E-12& 0.355804E-13\\
\hline
\end{tabular}}
\caption{Composite profiles for h $<$ 1 km after the {\it 'normalization'} for the f$_{gl}$ factor and the dome contribution included. These composite profiles are statistically representative for 43 nights. In the first column the height, in the second, third and fourth columns the J values for the 'good', 'typical' and 'bad caes. In the last line are reported the J values without the dome contribution. (median values: $\varepsilon_{d,25}$ = 0.35 arcsec, $\varepsilon_{d,50}$ = 0.52 arcsec, $\varepsilon_{d,75}$ = 0.70 arcsec). The first grid point (h = 0) includes $\CN2$ values between -12.5 m and +12.5 m. }
\label{comp_h_lt_1km_hvr}
\end{table*}

\section*{Acknowledgments}
This study has been funded by the Marie Curie Excellence Grant (FOROT) - MEXT-CT-2005-023878. 
We sincerely thank the VATT and LBT staff for the support offered during the Generalized Scidar runs at Mt. Graham. ECMWF products are extracted by the MARS catalog $http://www.ecmwf.int$ and authors are authorized to use them by the Meteorologic Service of the Italian Air Force. We thank R. Boyle and S. Egner for training J. Stoesz to guide the VATT telescope.



\label{lastpage}
\end{document}